\documentclass[12pt]{article}
\pdfoutput=1

\usepackage{booktabs}
\usepackage{tabulary}
\usepackage{multirow}
\usepackage{rotating}
\usepackage{epsfig}
\usepackage{psfrag}
\usepackage{latexsym}
\usepackage{indentfirst}
\usepackage{fancyhdr}
\usepackage{dsfont}
\usepackage{amsmath}
\usepackage{amssymb}
\usepackage{amsfonts}
\usepackage{mathrsfs}
\usepackage{amsthm}
\usepackage{pifont}
\usepackage{dsfont}
\usepackage{multirow}
\usepackage{array}
\usepackage{chngpage}
\usepackage{longtable}
\usepackage{cite}
\usepackage{bbold}
\usepackage{color}
\usepackage{braket}
\usepackage{colordvi}
\usepackage{fancybox}
\usepackage[footnotesize]{caption2}
\usepackage{graphicx}
\usepackage[center,footnotesize,hang]{subfigure}
\usepackage{bbm}
\usepackage{url}
\usepackage[colorlinks, linkcolor=red, anchorcolor=black, citecolor=green]{hyperref}
\usepackage{multirow}
\usepackage{fancybox}
\usepackage{harpoon}
\usepackage{textcomp}
\usepackage{enumitem}
\usepackage{mathrsfs}
\newcommand{\PreserveBackslash}[1]{\let\temp=\\#1\let\\=\temp}
\newcolumntype{C}[1]{>{\PreserveBackslash\centering}p{#1}}
\newcolumntype{R}[1]{>{\PreserveBackslash\raggedleft}p{#1}}
\newcolumntype{L}[1]{>{\PreserveBackslash\raggedright}p{#1}}
\allowdisplaybreaks

\newcommand{\bq}{\begin{eqnarray}}
\newcommand{\nq}{\end{eqnarray}}

\begin{document}
\title{
\begin{flushright}
\hfill\mbox{\small USTC-ICTS-19-26} \\[5mm]
\begin{minipage}{0.2\linewidth}
\normalsize
\end{minipage}
\end{flushright}
{\Large \bf
Modular $S_4$ and $A_4$ Symmetries and Their Fixed Points: New Predictive Examples of Lepton Mixing
\\[2mm]}}
\date{}

\author{
Gui-Jun~Ding$^{1}$\footnote{E-mail: {\tt
dinggj@ustc.edu.cn}},  \
Stephen~F.~King$^{2}$\footnote{E-mail: {\tt king@soton.ac.uk}}, \
Xiang-Gan Liu$^{1}$\footnote{E-mail: {\tt
hepliuxg@mail.ustc.edu.cn}},\
Jun-Nan Lu$^{1}$\footnote{E-mail: {\tt
hitman@mail.ustc.edu.cn}}  \
\\*[20pt]
\centerline{
\begin{minipage}{\linewidth}
\begin{center}
$^1${\it \small
Interdisciplinary Center for Theoretical Study and  Department of Modern Physics,\\
University of Science and Technology of China, Hefei, Anhui 230026, China}\\[2mm]
$^2${\it \small
Physics and Astronomy,
University of Southampton,
Southampton, SO17 1BJ, U.K.}\\
\end{center}
\end{minipage}}
\\[10mm]}
\maketitle
\thispagestyle{empty}

\begin{abstract}
In the modular symmetry approach to neutrino models, the flavour symmetry emerges as a finite subgroup $\Gamma_N$ of the modular symmetry, broken by the vacuum expectation value (VEV) of a modulus field $\tau$.
If the VEV of the modulus $\tau$ takes some special value, a residual subgroup of $\Gamma_N$ would be preserved.
We derive the fixed points $\tau_S=i$, $\tau_{ST}=(-1+i\sqrt{3})/2$, $\tau_{TS}=(1+i\sqrt{3})/2$, $\tau_T=i\infty$ in the fundamental domain which are invariant under the modular transformations indicated.  We then generalise these fixed points to
$\tau_f=\gamma\tau_S$, $\gamma\tau_{ST}$, $\gamma\tau_{TS}$ and $\gamma\tau_{T}$ in the upper half complex plane, and show that it is sufficient to consider $\gamma\in\Gamma_{N}$.
Focussing on level $N=4$, corresponding to the flavour group $S_4$,
we consider all the resulting triplet modular forms at these fixed points up to weight 6.
We then apply the results to lepton mixing, with different residual subgroups in the charged lepton sector and each of the right-handed neutrinos sectors. In the minimal case of two right-handed neutrinos,
we find three phenomenologically viable cases in which the light neutrino mass matrix only depends on three free parameters, and the lepton mixing takes the trimaximal TM1 pattern for two examples. One of these cases corresponds to a new Littlest Modular Seesaw based on CSD$(n)$ with $n=1+\sqrt{6}\approx 3.45$, intermediate between CSD$(3)$ and CSD$(4)$.
Finally, we generalize the results to examples with three right-handed neutrinos, also considering the level $N=3$ case, corresponding to $A_4$ flavour symmetry.

\end{abstract}
\newpage

%%%%%%%%%%%%%%%%%%%%%%%%%%%%%%%%%%%%%%%%%%%%%%%%%%%%%%%
\section{\label{sec:introduction}Introduction}
It is well-known that there are huge mass hierarchies among the quarks and leptons, and the quark mixing angles are small while the lepton sector has two large mixing angles $\theta_{12}$, $\theta_{23}$ and one small mixing angles $\theta_{13}$ which is of the same order of magnitude as the quark Cabibbo mixing angle~\cite{Tanabashi:2018oca}. The origin of the flavour structure of the quarks and leptons such as the mass hierarchies, mixing angles and CP violation phases is a big mystery of particle physics. Considerable effort has been devoted to understanding fermion masses and flavour mixing from flavour symmetry for decades. In particular, it is found that the observed lepton mixing angles can be naturally explained by the non-abelian discrete flavour symmetry, see Refs.~\cite{Altarelli:2010gt,Ishimori:2010au,King:2013eh,King:2014nza,King:2015aea,King:2017guk} for review. Moreover, the leptonic CP violation phases can be predicted and the precisely measured quark CKM mixing matrix can be accommodated if the discrete flavour symmetry is combined with generalized CP symmetry~\cite{Lu:2016jit,Li:2017abz,Lu:2018oxc,Lu:2019gqp}.

In the usual paradigm of discrete flavour symmetry, the standard model gauge symmetry is extended by certain finite flavour symmetry at high energy scale which is subsequently broken down to different subgroups in the neutrino and charged lepton sectors at low energy. In general some flavon fields which are standard model singlets are required to realize the flavour symmetry breaking. The flavon fields usually obtain vacuum expectation values (VEVs) along specific directions in order to reproduce phenomenologically viable lepton mixing angles. As a consequence, the scalar potential of discrete flavour symmetry models is rather elaborate, and certain auxiliary abelian symmetries are usually needed to forbid dangerous operators.

Recently, modular symmetry has been suggested as the origin of flavour symmetry~\cite{Feruglio:2017spp}. In this new framework, flavon fields might not be needed and the flavour symmetry can be uniquely broken by the VEV of the modulus $\tau$. Moreover, all higher-dimensional operators in the superpotential are completely determined by modular invariance if supersymmetry is exact. In modular invariant models, the Yukawa couplings transform nontrivially under the modular symmetry and they are modular forms which are holomorphic functions of $\tau$~\cite{Feruglio:2017spp}. Models with modular flavour symmetry can be highly predictive; the neutrino masses and mixing parameters can be predicted in terms of few input parameters, although the predictive power of this framework may be reduced by the K$\ddot{\mathrm{a}}$hler potential which is less  constrained by modular symmetry~\cite{Chen:2019ewa}.

The finite modular groups $\Gamma_2\cong S_3$~\cite{Kobayashi:2018vbk,Kobayashi:2018wkl,Kobayashi:2019rzp,Okada:2019xqk}, $\Gamma_3\cong A_4$~\cite{Feruglio:2017spp,Criado:2018thu,Kobayashi:2018vbk,Kobayashi:2018scp,Okada:2018yrn,Kobayashi:2018wkl,Novichkov:2018yse,Nomura:2019yft,Ding:2019zxk}, $\Gamma_4\cong S_4$~\cite{Penedo:2018nmg,Novichkov:2018ovf,Kobayashi:2019mna} and $\Gamma_5\cong A_5$~\cite{Novichkov:2018nkm,Ding:2019xna} have been studied and some simple modular models have been constructed. It is remarkable that even the $A_4$ modular models can reproduce the measured neutrino masses and mixing angles\cite{Feruglio:2017spp,Kobayashi:2018scp,Ding:2019zxk}. The modular invariance approach has been extended to include odd weight modular forms which can be decomposed into irreducible representations of the the homogeneous finite modular group $\Gamma'_{N}$~\cite{Liu:2019khw}, and the modular symmetry $\Gamma'_{3}\cong T'$ has been discussed. It has been shown that the modular symmetry can be consistently combined with generalized CP symmetry, and the modulus transform as $\tau\rightarrow-\tau^{*}$ under the CP transformation~\cite{Novichkov:2019sqv,Baur:2019kwi,Acharya:1995ag,Dent:2001cc,Giedt:2002ns}. Motivated by factorized tori compactification in superstring theory, the formalism of the single modulus has been generalized to the case of a direct product of multiple moduli~\cite{deMedeirosVarzielas:2019cyj,King:2019vhv}. From the view of top-down, the modular symmetry naturally appears in string constructions~\cite{Kobayashi:2018rad,Kobayashi:2018bff,Baur:2019kwi,Baur:2019iai}.

It has been realised that, if the VEV of the modulus $\tau$ takes some special value, a residual subgroup of the finite modular symmetry group $\Gamma_N$ would be preserved. The phenomenological implications of the residual modular symmetry have been discussed in the context of modular $A_4$~\cite{Novichkov:2018yse}, $S_4$~\cite{Novichkov:2018ovf} and $A_5$~\cite{Novichkov:2018nkm} symmetries. If the modular symmetry is broken down to a residual $Z_3$ (or $Z_5$) subgroup in charged lepton sector and to a $Z_2$ subgroup in the neutrino sector, the trimaximal TM1 and TM2 mixing patterns can be obtained~\cite{Novichkov:2018yse,Novichkov:2018ovf}.

In this paper, we derive the fixed points $\tau_S=i$, $\tau_{ST}=(-1+i\sqrt{3})/2$, $\tau_{TS}=(1+i\sqrt{3})/2$, $\tau_T=i\infty$ in the fundamental domain which are invariant under the modular transformations indicated.  We then generalise these fixed points to
$\tau_f=\gamma\tau_S$, $\gamma\tau_{ST}$, $\gamma\tau_{TS}$ and $\gamma\tau_{T}$ in the upper half complex plane, and show that it is sufficient to consider $\gamma\in\Gamma_{N}$.
Focussing on level $N=4$, corresponding to the flavour group $S_4$,
we consider all the resulting triplet modular forms at these fixed points up to weight 6.
We then apply the results to obtain new predictive examples of lepton mixing, with different residual subgroups in the charged lepton sector and each of the right-handed neutrinos sectors.
In the minimal case of two right-handed neutrinos,
we find three phenomenologically viable cases in which the light neutrino mass matrix only depends on three free parameters, and the lepton mixing takes the trimaximal $\mathrm{TM}_1$ pattern for two examples.
Finally, we generalize the results to examples with three right-handed neutrinos, also considering the level $N=3$ case, corresponding to $A_4$ flavour symmetry, listing the values of modular forms at the fixed points in this case also.

It is interesting to compare the modular symmetry approach here to the
tri-direct CP approach in usual discrete flavour symmetry based on the residual symmetry~\cite{Ding:2018fyz,Ding:2018tuj},
which can also give rise to very predictive models such as the Littlest seesaw model based on CSD$(n)$ with $n=3,4$ or a variant~\cite{King:2013iva,King:2013xba,King:2015dvf,Chen:2019oey}.
In the present work, we shall also follow the tri-direct approach but in the context of modular invariant models and without CP. In the minimal scenario, with two right-handed neutrinos, the two columns of the Dirac neutrino mass matrix are modular forms which are aligned along certain directions at some special values of $\tau$.
One of the minimal $S_4$ cases corresponds to a new Littlest Modular Seesaw based on
CSD$(n)$ with $n=1+\sqrt{6}\approx 3.45$, intermediate between CSD$(3)$ and CSD$(4)$.

The paper is organized as follows. In section~\ref{sec:modularform_of_N=4}, we give a brief review on the modular symmetry and the modular forms of level 4 are constructed from the products of the Dedekind $\eta$ function. In section~\ref{sec:fp&ResSym}, we analyze the nontrivial fixed points $\tau_{f}$ of the complex modulus which preserves a residual modular subgroup. We find there are infinity fixed modulus with the form $\tau_f=\gamma\tau_S$, $\gamma\tau_{ST}$, $\gamma\tau_{TS}$ and $\gamma\tau_{T}$, where $\gamma$ is an arbitrary modular transformation and $\tau_S=i$, $\tau_{ST}=-\frac{1}{2}+i\frac{\sqrt{3}}{2}$, $\tau_{TS}=\frac{1}{2}+i\frac{\sqrt{3}}{2}$, $\tau_T=i\infty$.
However, the independent alignments of modular forms at the fixed points are finite. In section~\ref{sec:tridirec_MM}, we present the framework of tri-direct modular model, the minimal scenario is based on the two right-handed neutrinos model, and we generalize this approach to the three right-handed neutrinos case. In order to show concrete examples, we analyze the tri-direct modular models for the $S_4$ group. Furthermore, in section~\ref{sec:Tri-direct_MM_A4} we present the cases which can also be obtained, if the modular symmetry is $A_4$ instead of $S_4$. We conclude in section~\ref{sec:conclusion}. Finally the group theory of $S_4$ as well as Clebsch-Gordan coefficients in our basis are collected in Appendix~\ref{sec:S4_group_app}.

\section{\label{sec:modularform_of_N=4}Modular symmetry and modular form multiplets of level $N=4$}

The modular group $\overline{\Gamma}$ is the group of linear fraction transformations which acts on the complex modulus $\tau$ in the upper half complex plane as follow,
\begin{equation}
\tau\rightarrow\gamma\tau=\frac{a\tau+b}{c\tau+d},~~\text{with}~~a, b, c, d\in\mathbb{Z},~~ad-bc=1,~~~\Im\tau>0\,.
\end{equation}
We note that the map
\begin{equation}
\frac{a\tau+b}{c\tau+d}\mapsto\left(
\begin{array}{cc}
a  &  b  \\
c  &  d
\end{array}
\right)
\end{equation}
is an isomorphism from the modular group to the projective
matrix group $PSL(2, \mathbb{Z})\cong SL(2, \mathbb{Z})/\{\pm I\} $, where $SL(2, \mathbb{Z})$ is the group of two-by-two matrices with integer entries and determinant equal to one. It is obvious that
\begin{equation}
\frac{a\tau+b}{c\tau+d}~~~~\text{is~the~same~as~}~~~~\frac{-a\tau-b}{-c\tau-d}\,,
\end{equation}
therefore we identify
\begin{equation}
\label{eq:gamma_fgamma}\left(
\begin{array}{cc}
a  &  b  \\
c  &  d
\end{array}
\right)~~~~\text{is~the~same~as~}~~~~\left(
\begin{array}{cc}
-a  &  -b  \\
-c  & -d
\end{array}
\right)
\end{equation}
in matrix notation. The modular group $\overline{\Gamma}$ can be generated by two generators $S$ and $T$
\begin{equation}
S:\tau\mapsto -\frac{1}{\tau},~~~~\quad T: \tau\mapsto\tau+1\,,
\end{equation}
which are represented by the following two matrices of $PSL(2, \mathbb{Z})$,
\begin{equation}
S=\left(\begin{array}{cc}
0 & 1 \\
-1  & 0
\end{array}
\right),~~~\quad T=\left(\begin{array}{cc}
1 & 1 \\
0  & 1
\end{array}
\right)\,.
\end{equation}
We can check that the generators $S$ and $T$ obey the relations,
\begin{equation}
\label{eq:multiply_rules}S^2=(ST)^3=(TS)^3=1\,.
\end{equation}
The principal congruence subgroup of level $N$ is the subgroup
\begin{equation}
\Gamma(N)=\left\{\begin{pmatrix}
a  & b \\
c  & d
\end{pmatrix}\in SL(2, \mathbb{Z}),~ b=c=0\,(\mathrm{mod~N}), a=d=1\,(\mathrm{mod~N})
\right\}\,,
\end{equation}
which is an infinite normal subgroup of $SL(2, \mathbb{Z})$. It is easy to see that $T^{N}$ is an element of $\Gamma(N)$. The projective principal congruence subgroup is defined as $\overline{\Gamma}(N)=\Gamma(N)/\{\pm I \}$ for $N=1, 2$. For the values of $N\geq3$, we have $\overline{\Gamma}(N)=\Gamma(N)$ because
$\Gamma(N)$ doesn't contain the element $-I$. The quotient group $\Gamma_N\equiv\overline{\Gamma}/\overline{\Gamma}(N)$ is the finite modular group, and it can be obtained by further imposing the condition $T^{N}=1$ besides those in Eq.~\eqref{eq:multiply_rules}.

A crucial element of the modular invariance approach is the modular form $f(\tau)$ of weight $k$ and level $N$. The modular form $f(\tau)$ is a holomorphic function of the complex modulus $\tau$ and it is required to transform under the action of $\overline{\Gamma}(N)$ as follows,
\begin{equation}
f\left(\frac{a\tau+b}{c\tau+d}\right)=(c\tau+d)^kf(\tau)~~~\mathrm{for}~~\forall~\left(
\begin{array}{cc}
a  &  b  \\
c  &  d
\end{array}
\right)\in\overline{\Gamma}(N)\,.
\end{equation}
The modular forms of weight $k$ and level $N$ span a linear space of finite dimension. It is always possible to choose a basis in this linear space such that the modular forms can be arranged into some modular multiplets $f_{\mathbf{r}}\equiv\left(f_1(\tau), f_{2}(\tau),...\right)^{T}$ which transform as irreducible representation $\mathbf{r}$ of the finite modular group $\Gamma_N$ for even $k$~\cite{Feruglio:2017spp,Liu:2019khw}, i.e.
\begin{equation}
f_{\mathbf{r}}(\gamma\tau)=(c\tau+d)^k\rho_{\mathbf{r}}(\gamma)f_{\mathbf{r}}(\tau)~~~\mathrm{for}~~\forall~\gamma\in\overline{\Gamma}\,,
\end{equation}
where $\gamma$ is the representative element of the coset
$\gamma\overline{\Gamma}(N)$ in $\Gamma_N$, and $\rho_{\mathbf{r}}(\gamma)$ is the representation matrix of the element $\gamma$ in the irreducible representation $\mathbf{r}$.

\subsection{Modular forms of level 4}

The modular forms of level 4 has been constructed in~\cite{Penedo:2018nmg,Novichkov:2018ovf} in terms of $\eta'(\tau)/\eta(\tau)$, where $\eta(\tau)$ and $\eta'(\tau)$ are the Dedekind eta function and its derivative. In this section, we shall construct the modular forms of level 4 from the products of $\eta(\tau)$. The modular forms of weight $k$ and level $4$ form a linear space $\mathcal{M}_{k}(\Gamma(4))$ as follow~\cite{schultz2015notes},
\begin{equation}
\label{eq:Mk_Gamma4}\mathcal{M}_{k}(\Gamma(4))=\bigoplus_{a+b=2k,\,a,b\ge0} \mathbb{C} \frac{\eta^{2b-2a}(4\tau)\eta^{5a-b}(2\tau)}{\eta^{2a}(\tau)}\,,
\end{equation}
where $a$, $b$ and $k$ are positive integers, and the Dedekind eta function $\eta(\tau)$ is defined as~\cite{diamond2005first,Bruinier2008The,lang2012introduction},
\begin{equation}
\eta(\tau)\equiv q^{1/24}\prod_{n=1}^\infty \left(1-q^n \right),~\quad\text{with} \quad q=e^{2\pi i\tau}\,,
\end{equation}
which satisfies the following well-known identities
\begin{equation}
\eta(\tau+1)=e^{i \pi/12}\eta(\tau),~~~~\eta(-1/\tau)=\sqrt{-i \tau}~\eta(\tau)\,.
\end{equation}
We can read from Eq.~\eqref{eq:Mk_Gamma4} that the linear space $\mathcal{M}_{k}(\Gamma(4))$ has dimension $2k+1$. As shown in~\cite{Feruglio:2017spp,Liu:2019khw}, the even weight modular forms of level 4 can be decomposed into irreducible representations of the inhomogeneous finite modular groups $\Gamma_4\cong S_4$, while odd weight modular forms can be arranged into irreducible representations of the homogeneous finite modular groups $\Gamma'_4$ which is the double covering group of $S_4$. In the present work, we shall focus on the modular forms of even weights. For the weight 2 modular forms with $k=2$, Eq.~\eqref{eq:Mk_Gamma4} implies that without loss of generality the basis vectors of $\mathcal{M}_{2}(\Gamma(4))$ could be chosen to be
\begin{equation}
\frac{\eta^{2b-2a}(4\tau)\eta^{5a-b}(2\tau)}{\eta^{2a}(\tau)}\,,
\end{equation}
with $(a, b)=(0, 4), (1, 3), (2, 2), (3, 1)$ and $(4, 0)$. To be more specific, the basis vectors of $\mathcal{M}_{2}(\Gamma(4))$ are
\begin{equation}
\begin{aligned}
& e_1(\tau)=\frac{\eta^{8}(4\tau)}{\eta^{4}(2\tau)},~\quad e_{2}(\tau)=\frac{\eta^{4}(4\tau)\eta^{2}(2\tau)}{\eta^{2}(\tau)},~\quad e_3(\tau)=\frac{\eta^{8}(2\tau)}{\eta^{4}(\tau)},\\
& e_4(\tau)=\frac{\eta^{14}(2\tau)}{\eta^{4}(4\tau)\eta^{6}(\tau)}, ~\quad e_5(\tau)=\frac{\eta^{20}(2\tau)}{\eta^{8}(4\tau)\eta^{8}(\tau)}\,.
\end{aligned}
\end{equation}
The $q$-expansions of the above basis vectors $e_i$ are given by
\begin{eqnarray}
\nonumber e_1(\tau)&=& q+4q^3+6q^5+8q^7+13q^9+\dots\,, \\
\nonumber e_2(\tau)&=& q^{3/4}(1+2q+3q^2+6q^3+5q^4+6q^5+10q^6+8q^7+12q^8+14q^9+\dots)\,, \\
\nonumber e_3(\tau)&=& q^{1/2}(1+4q+6q^2+8q^3+13q^4+12q^5+14q^6+24q^7+18q^8+20q^9+ \dots)\,,\\
\nonumber e_4(\tau)&=& q^{1/4}(1+6q+13q^2+14q^3+18q^4+32q^5+31q^6+30q^7+48q^8+38q^9\dots)\,, \\
\label{eq:ei_expa}e_5(\tau)&=& 1+8q+24q^2+32q^3+24q^4+48q^5+96q^6+64q^7+24q^8+104q^9+\dots\,.
\label{eq:q_exp_ei}
\end{eqnarray}
Note that any modular form of weight 2 and level $4$ can be written as a linear combination of $e_{1,2,3,4,5}$. Under the action of the generator $T$, it is easy to check that the basis vectors $e_i$ transform as
\begin{eqnarray}
e_1(\tau)\stackrel{T}{\longmapsto}e_1(\tau),\quad e_{2}(\tau)\stackrel{T}{\longmapsto} -ie_2 ,\quad e_{3}(\tau)\stackrel{T}{\longmapsto} -e_3\,,\quad e_{4}(\tau)\stackrel{T}{\longmapsto} ie_4\,,\quad e_{5}(\tau)\stackrel{T}{\longmapsto} e_5\,.
\end{eqnarray}
Under another generator $S$, we have
\begin{eqnarray}
\nonumber&& e_1(\tau) \stackrel{S}{\longmapsto} -\frac{\tau^2}{64} \left(16e_1-32e_2+24e_3-8e_4+e_5 \right), \\
\nonumber&& e_2(\tau)\stackrel{S}{\longmapsto} -\frac{\tau^2}{32}\left(-16e_1+16e_2-4e_4+e_5 \right), \\
\nonumber&& e_3(\tau)\stackrel{S}{\longmapsto} -\frac{\tau^2}{16} \left(16e_1-8e_3+e_5 \right), \\
\nonumber&& e_4(\tau)\stackrel{S}{\longmapsto} -\frac{\tau^2}{8} \left(-16e_1-16e_2+4e_4+e_5 \right), \\
&& e_5(\tau) \stackrel{S}{\longmapsto}-\frac{\tau^2}{4} \left(16e_1+32e_2+24e_3+8e_4+e_5 \right)\,.
\end{eqnarray}
We see that the basis vectors $e_{1,2,3,4,5}$ are closed under $S$ and $T$ up to multiplicative factors, and they are mapped into themselves by the elements $S^2$, $(ST)^3$, $(TS)^3$ and $T^4$. The above five modular forms can be organized into a doublet $\mathbf{2}$ and a triplet $\mathbf{3}$ of the finite modular group $S_4$\footnote{In our working basis, the triplet representations $\mathbf{3}$ and $\mathbf{3}'$ correspond to $\mathbf{3}'$ and $\mathbf{3}$ of~\cite{Penedo:2018nmg,Novichkov:2018ovf} respectively. },
\begin{equation}
\label{eq:modular_space}
Y^{(2)}_{\mathbf{2}}(\tau)=\begin{pmatrix}
Y_1(\tau) \\
Y_2(\tau)
\end{pmatrix}\,,
\quad
Y^{(2)}_{\mathbf{3}}(\tau)=\begin{pmatrix}
Y_3(\tau) \\
Y_4(\tau) \\
Y_5(\tau)
\end{pmatrix}\,,
\end{equation}
where
\begin{align}
\nonumber
Y_1(\tau)&=16\omega^2e_1(\tau)-8(2+\omega^2)e_3(\tau)+\omega^2e_5(\tau),\\
\nonumber
Y_2(\tau)&=16e_1(\tau)+8i\sqrt{3}e_3(\tau)+e_5(\tau), \\
\nonumber
Y_3(\tau)&=-\omega^2\left[16e_1(\tau)+16(1-i)e_2(\tau)+4(1+i)e_4(\tau) -e_5(\tau)\right], \\
\nonumber
Y_4(\tau)&=-\omega\left[16e_1(\tau)+8(1-\sqrt{3})(-1+i)e_2(\tau)-2(1+\sqrt{3})(1+i)e_4(\tau)-e_5(\tau)\right],\\
Y_5(\tau)&=-16e_1(\tau)+8(1+\sqrt{3})(1-i)e_2(\tau)+2(1-\sqrt{3})(1+i)e_4(\tau)+e_5(\tau)\,,
\end{align}
with $\omega=e^{i2\pi/3}$. From Eq.~\eqref{eq:ei_expa}, we can read out the $q$-expansion of $Y^{(2)}_{\mathbf{2}}$ and $Y^{(2)}_{\mathbf{3}}$ as follows,
\begin{eqnarray}
\label{eq:q_exp_weight2}
\nonumber Y_1(\tau)&=&\omega^2\left(1 + 24 q_1 -72 q_1^2 -288 q_1^3 + 216 q_1^4 + 1296 q_1^5 - 2592 q^6 -5184q_1^7 + \dots\right)\,,\\
\nonumber Y_2(\tau)&=& 1 - 24 q_1 -72 q_1^2 + 288 q_1^3 + 216 q_1^4 - 1296 q_1^5 - 2592 q^6 + 5184q_1^7 +\dots\,, \\
\nonumber Y_3(\tau)&=& \omega^2(1- 8q_2 + 64q_2^3 + 32q_2^4 + 192q_2^5 - 512q_2^7 + 384q_2^8 - 1664q_2^9 +\dots)\,,\\
\nonumber Y_4(\tau)&=&\omega(1 - 8a q_2+ 64bq_2^3 + 32 q_2^4 + 192a q_2^5 -512bq_2^7 +384q_2^8 -1664a q_2^9 +\dots)\,, \\
 Y_5(\tau)&=& 1 - 8b q_2+ 64aq_2^3 + 32 q_2^4 + 192b q_2^5 -512aq_2^7 +384q_2^8 -1664b q_2^9 +\dots\,,
\end{eqnarray}
where the constants $a=(-1-\sqrt{3})/2$, $b=(-1+\sqrt{3})/2$, $q_1=-i\sqrt{q/3}$ and $q_2=\dfrac{1}{2}(1+i)q^{1/4}$. Form the above expressions of $q$-expansion, we see that the modular forms $Y_i\,(i=1,\ldots,5)$ satisfy the following constraints:
\begin{eqnarray}
\nonumber&&3Y_1^2-\omega(Y_5^2+2Y_3Y_4)=0,~~\quad \sqrt{3}\left(Y_2Y_3-Y_1Y_5\right)-\omega^2(Y_5^2-Y_3Y_4)=0\,,\\
\nonumber&&3Y_2^2-\omega(Y_4^2+2Y_3Y_5)=0,~~\quad \sqrt{3}\left(Y_2Y_4-Y_1Y_3\right)-\omega^2(Y_4^2-Y_3Y_5)=0\,,\\
&&3Y_1Y_2-\omega(Y_3^2+2Y_4Y_5)=0,~~\quad \sqrt{3}\left(Y_2Y_5-Y_1Y_4\right)-\omega^2(Y_3^2-Y_4Y_5)=0\,.
\end{eqnarray}
The weight 4 modular forms can be generated from the tensor products of $Y^{(2)}_{\mathbf{2}}$ and $Y^{(2)}_{\mathbf{3}}$.
There are 9 linearly independent weight 4 modular forms which can be arranged into $S_4$ irreducible representations $\mathbf{1}$, $\mathbf{1}'$, $\mathbf{2}$, $\mathbf{3}$, $\mathbf{3}'$ as follow:
\begin{eqnarray}
\nonumber&& Y^{(4)}_{\mathbf{1},I}=\left(Y^{(2)}_{\mathbf{2}}Y^{(2)}_{\mathbf{2}}\right)_{\mathbf{1}}= 2Y_1Y_2 , \\
\nonumber&& Y^{(4)}_{\mathbf{2},I}=\left(Y^{(2)}_{\mathbf{2}}Y^{(2)}_{\mathbf{2}}\right)_{\mathbf{2}}=\Big( Y^2_2, Y^2_1 \Big)^T, \\
\nonumber&& Y^{(4)}_{\mathbf{3},I}=\left(Y^{(2)}_{\mathbf{2}}Y^{(2)}_{\mathbf{3}}\right)_{\mathbf{3}}=\Big(Y_1Y_4+Y_2Y_5, Y_2Y_3+Y_1Y_5, Y_1Y_3+Y_2Y_4 \Big)^T, \\
&& Y^{(4)}_{\mathbf{3'},I}=\left(Y^{(2)}_{\mathbf{2}}Y^{(2)}_{\mathbf{3}}\right)_{\mathbf{3'}}=\Big(Y_1Y_4-Y_2Y_5, -Y_2Y_3+Y_1Y_5, Y_1Y_3-Y_2Y_4 \Big)^T\,.
\end{eqnarray}
Notice that the other possible modular multiplets are either vanishing or parallel to the above modular multiplets:
\begin{eqnarray}
\nonumber&&Y^{(4)}_{\mathbf{1'}}=\left(Y^{(2)}_{\mathbf{2}}Y^{(2)}_{\mathbf{2}}\right)_{\mathbf{1'}}=0,~~~\quad Y^{(4)}_{\mathbf{3},II}=\left(Y^{(2)}_{\mathbf{3}}Y^{(2)}_{\mathbf{3}}\right)_{\mathbf{3}}=(0,\,0,\,0)^T\,,\\
\nonumber&& Y^{(4)}_{\mathbf{1},II} = \left(Y^{(2)}_{\mathbf{3}}Y^{(2)}_{\mathbf{3}}\right)_{\mathbf{1}} =\frac{3}{2\omega} Y^{(4)}_{\mathbf{1},I},~~~\quad Y^{(4)}_{\mathbf{2}, II} = \left(Y^{(2)}_{\mathbf{3}}Y^{(2)}_{\mathbf{3}}\right)_{\mathbf{2}} =\frac{3}{\omega} Y^{(4)}_{\mathbf{2}, I}, \\
&& Y^{(4)}_{\mathbf{3'},II} = \left(Y^{(2)}_{\mathbf{3}}Y^{(2)}_{\mathbf{3}}\right)_{\mathbf{3'}} =-2\sqrt{3}\omega Y^{(4)}_{\mathbf{3'},I}\,.
\end{eqnarray}
The linear space of modular forms of level 4 and weight 6 has dimension $2k+1=2\times6+1=13$, under the finite modular group $S_4$ it can be decomposed into
\begin{eqnarray}
\nonumber&&Y^{(6)}_{\mathbf{1}}=\left(Y^{(2)}_{\mathbf{2}}Y^{(4)}_{\mathbf{2},I}\right)_{\mathbf{1}}= Y^3_1+Y^3_2 , \\
\nonumber&&Y^{(6)}_{\mathbf{1'}}=\left(Y^{(2)}_{\mathbf{2}}Y^{(4)}_{\mathbf{2},I}\right)_{\mathbf{1'}}= Y^3_1-Y^3_2, \\
\nonumber&&Y^{(6)}_{\mathbf{2}}=\left(Y^{(2)}_{\mathbf{2}}Y^{(4)}_{\mathbf{1},I}\right)_{\mathbf{2}}=\Big(2Y^2_1Y_2, 2Y_1Y^2_2\Big)^T, \\
\nonumber&&Y^{(6)}_{\mathbf{3}, I}=\left(Y^{(2)}_{\mathbf{3}}Y^{(4)}_{\mathbf{1},I}\right)_{\mathbf{3}}=\Big(2 Y_1Y_2Y_3, 2Y_1Y_2Y_4, 2Y_1Y_2Y_5\Big)^T, \\
\nonumber&&Y^{(6)}_{\mathbf{3}, II}=\left(Y^{(2)}_{\mathbf{3}}Y^{(4)}_{\mathbf{2},I}\right)_{\mathbf{3}}=\Big(Y_2^2Y_4 + Y_1^2Y_5, Y_1^2Y_3 + Y_2^2Y_5, Y_2^2Y_3 + Y_1^2 Y_4 \Big)^T\\
&&Y^{(6)}_{\mathbf{3'}}=\left(Y^{(2)}_{\mathbf{3}}Y^{(4)}_{\mathbf{2},I}\right)_{\mathbf{3'}}=\Big(Y^2_2Y_4-Y^2_1Y_5,-Y^2_1Y_3+Y^2_2Y_5, Y^2_2Y_3-Y^2_1Y_4 \Big)^T\,.
\end{eqnarray}
Higher weight modular forms can be constructed in the same fashion, see Refs.~\cite{Penedo:2018nmg,Novichkov:2018ovf}  for modular forms of weight 8 and weight 10. Note that in our working basis the representation matrices are different from those of~\cite{Penedo:2018nmg,Novichkov:2018ovf}, and they are related to the choices of~\cite{Penedo:2018nmg,Novichkov:2018ovf} by unitary transformation.

%%%%%%%%%%%%%%%%%%%%%%%%%%%%%%%%%%%%%%%%%%%%%%%%%%%%%%%
\section{\label{sec:fp&ResSym}Fixed points and residual modular symmetry  }

In this section, we shall first discuss the fixed point in the fundamental domain, subsequently we study all the possible fixed points in the upper half complex plane. Finally we investigate the constraints on the neutrino and charged lepton mass matrices imposed by the residual modular symmetries at fixed points.

\subsection{Fixed point in the fundamental domain}

If a modulus parameter $\tau_0$ is invariant under the action of a nontrivial $SL(2, \mathbb{Z})$ transformation $\gamma_0\neq\pm I$, we call $\tau_0$ is the fixed point of $\gamma_0$ and $\gamma_0$ is the stabilizer of $\tau_0$, i.e.
\begin{equation}
\label{eq:fp_equa}\gamma_0\tau_0=\tau_0\,.
\end{equation}
The fundamental domain of the modular group is denoted as $\mathcal{F}: \left|\Re\tau\right|\leq\frac{1}{2}, \Im\tau>0, \left|\tau\right|\geq1$. Firstly we consider the case that the fixed point $\tau_0$ is in the fundamental region $\tau_0\in\mathcal{F}$. The modular transformation $\gamma_0$ can be generally parameterized as
\begin{equation}
\gamma_{0}=\left(\begin{array}{cc}
a_0 & b_0 \\
c_0 & d_0
\end{array}\right)\,,\quad \text{with} ~~~
a_0, b_0, c_0, d_0\in \mathbb{Z}~~\text{and}~~a_0d_0-b_0c_0=1\,.
\end{equation}
Thus the fixed point condition of Eq.~\eqref{eq:fp_equa} gives
\begin{equation}
\label{eq:fp_condition}\frac{a_0\tau_{0}+b_0}{c_0\tau_{0}+d_0}=\tau_{0}\,.
\end{equation}
Multiplying $c_0\tau_{0}+d_0$ on both sides of Eq.~\eqref{eq:fp_condition}, we can obtain\footnote{Notice that $c_0\tau_{0}+d_0\neq 0$ for $\tau_0$ in the fundamental domain.}
\begin{equation}
\label{eq:fixed_eq}
c_0\tau_{0}^{2}+(d_0-a_0)\tau_{0}-b_0=0\,.
\end{equation}
If $c_0=0$, then we have $\tau_{0}=\frac{b_0}{d_0-a_0}$ which is outside the fundamental domain $\mathcal{F}$. Consequently $c_0$ should be non-vanishing with $c_0\neq 0$. Thus the roots of equation in Eq.~\eqref{eq:fixed_eq} are given by
\begin{equation}
\label{eq:roots}\tau_0=\frac{a_0-d_0\pm \sqrt{(a_0+d_0)^{2}-4}}{2c_0}\,.
\end{equation}
Since the modular parameter $\tau_{0}$ is in the fundamental domain and its imaginary part is positive $\Im\tau_{0}>0$, the constraint $\left|a_0+d_0\right|<2$ should be satisfied. Therefore we have $a_0+d_0=0, \pm1$, as the parameters $a_0$, $b_0$, $c_0$ and $d_0$ are all integers.
\begin{itemize}
\item{$a_0+d_0=0$}

In this case, we have $a_0=-d_0$. The fixed modulus $\tau_0$ in Eq.~\eqref{eq:roots} is of the following form,
\begin{equation}
\tau_{0}=\frac{a_0}{c_0}\pm\frac{i}{c_0}\,.
\end{equation}
Since $\tau_{0}$ is in fundamental region with $|\Re\tau_{0}|\leq\frac{1}{2}$ and $\left|\tau_{0}\right|\geq1$, thus we have the following constraint,
\begin{equation}
\left|\frac{a_0}{c_0}\right|\leq \frac{1}{2},~\quad~ \left|\frac{1}{c_0}\right|\geq \frac{\sqrt{3}}{2}\,,
\end{equation}
which implies
\begin{equation}
a_0=d_0=0,~\quad~c_0=\pm1\,.
\end{equation}
Moreover, the unit determinant of $\gamma_{0}$ requires $a_0d_0-b_0c_0=1$, thus we find
\begin{equation}
b_0=-c_0=\mp1\,.
\end{equation}
Hence the modular transformation $\gamma_0$ takes the form
\begin{equation}
\gamma_{0}=\mp\left( \begin{array}{cc} 0 & 1 \\ -1 & 0  \end{array} \right)=\mp S\,.
\end{equation}
Notice that $S$ and $-S$ lead to the same linear fractional transformation, as shown in Eq.~\eqref{eq:gamma_fgamma}. Accordingly the fixed point is $\tau_0=i\equiv\tau_S$.

\item{$a_0+d_0=1$}

The parameter $d_0$ can be expressed in terms of $a_0$ as
\begin{equation}
d_0=1-a_0\,.
\end{equation}
The solution for $\tau_0$ in Eq.~\eqref{eq:roots} is simplified into
\begin{equation}
\tau_{0}=\frac{2a_0-1\pm i\sqrt{3}}{2c_0}\,.
\end{equation}
The requirement of $\tau_0\in\mathcal{F}$ entails
\begin{equation}
\left|\frac{2a_0-1}{2c_0}\right|\leq \frac{1}{2}\,,~\quad~ \left|\frac{\sqrt{3}}{2c_0}\right|\geq \frac{\sqrt{3}}{2}\,,
\end{equation}
which leads to
\begin{equation}
a_0=0,~~b_0=\mp1,~~c_0=\pm1,~~d_0=1\,,
\end{equation}
or
\begin{equation}
a_0=1,~~b_0=\mp1,~~c_0=\pm1,~~d_0=0\,.
\end{equation}
As a consequence, $\gamma_0$ and $\tau_0$ are fixed to be
\begin{equation}
\begin{aligned}
 &\gamma_{0}=\left( \begin{array}{cc} 0 & -1 \\ 1 & 1 \end{array} \right)=-ST, ~~~~\tau_0=\frac{-1+i\sqrt{3}}{2}\,,\\
&\gamma_{0}=\left( \begin{array}{cc} 0 & 1 \\ -1 & 1 \end{array} \right)=-(TS)^{2},~~~~\tau_0=\frac{1+i\sqrt{3}}{2}\,,
\end{aligned}
\end{equation}
or
\begin{equation}
\begin{aligned}
&\gamma_{0}=\left( \begin{array}{cc} 1 & -1 \\ 1 & 0 \end{array} \right)=-TS,~~~~\tau_0=\frac{1+i\sqrt{3}}{2}\,,\\
&\gamma_{0}=\left( \begin{array}{cc} 1 & 1 \\ -1 & 0 \end{array} \right)=-(ST)^{2},~~~~\tau_0=\frac{-1+i\sqrt{3}}{2}\,.
\end{aligned}
\end{equation}

\item{$a_0+d_0=-1$}

In this case, the parameter $d_0=-1-a_0$ and the solution of $\tau_0$ is
\begin{equation}
\tau_{0}=\frac{2a_0+1\pm i\sqrt{3}}{2c_0}\,.
\end{equation}
The condition $\tau_0\in\mathcal{F}$ imposes the following constraints,
\begin{equation}
\left|\frac{2a_0+1}{2c_0}\right|\leq \frac{1}{2}\,,~\quad~ \left|\frac{\sqrt{3}}{2c_0}\right|\geq \frac{\sqrt{3}}{2}\,.
\end{equation}
Hence the parameters $a_0$, $b_0$, $c_0$ and $d_0$ are determined to be
\begin{equation}
a_0=-1,~~~b_0=\mp1,~~~c_0=\pm1,~~~d_0=0\,,
\end{equation}
or
\begin{equation}
a_0=0,~~~b_0=\mp1,~~~c_0=\pm1,~~~d_0=-1\,.
\end{equation}
The modular transformation $\gamma_0$ and the corresponding fixed point $\tau_0$ are found to be given by
\begin{equation}
\begin{aligned}
&\gamma_{0}=\left( \begin{array}{cc} -1 & -1 \\ 1 & 0 \end{array} \right)=(ST)^{2},~~~\tau_0=\frac{-1+i\sqrt{3}}{2}\,,\\
&\gamma_{0}=\left( \begin{array}{cc} -1 & 1 \\ -1 & 0 \end{array} \right)=TS,~~~\tau_0=\frac{1+i\sqrt{3}}{2}\,,
\end{aligned}
\end{equation}
or
\begin{equation}
\begin{aligned}
&\gamma_{0}=\left( \begin{array}{cc} 0 & -1 \\ 1 & -1 \end{array} \right)=(TS)^{2},~~~\tau_0=\frac{1+i\sqrt{3}}{2}\,,\\
&\gamma_{0}=\left( \begin{array}{cc} 0 & 1 \\ -1 & -1 \end{array} \right)=ST,~~~\tau_0=\frac{-1+i\sqrt{3}}{2}\,.
\end{aligned}
\end{equation}
We see that $\tau_{ST}=\frac{-1+i\sqrt{3}}{2}$ and $\tau_{TS}=\frac{1+i\sqrt{3}}{2}$ are the fixed points of the modular transformation $\pm ST$ and $\pm TS$ respectively. Moreover they are related by modular transformation $T$,
\begin{equation}
T\tau_{ST}=\tau_{TS}\,.
\end{equation}
Furthermore, there is a fourth fixed point $\tau_{0}=i\infty$. It is easy to check that $i\infty$ is invariant under the action of $T$ as
$T\left(i\infty\right)=i\infty+1=i\infty$. We shall denote $\tau_{T}\equiv i\infty$ in the following. In short, we have only the following four nontrivial fixed points in the fundamental domain,
\begin{equation}
\tau_S=i,~~~\tau_{ST}=-\frac{1}{2}+i\frac{\sqrt{3}}{2},~~~ \tau_{TS}=\frac{1}{2}+i\frac{\sqrt{3}}{2},~~~ \tau_T=i\infty\,.
\end{equation}

\end{itemize}

In general, a modular multiplet $Y^{(k)}_{\mathbf{r}}(\tau)$ of weight $k$ in the irreducible representation $\mathbf{r}$ of the finite modular group $\Gamma_N$ satisfies the following property,
\begin{equation}
\label{eq:MF_irr}Y^{(k)}_{\mathbf{r}}(\gamma\tau)=J_k(\gamma, \tau)\rho_{\mathbf{r}}(\gamma)Y^{(k)}_{\mathbf{r}}(\tau)\,,
\end{equation}
where $J_k(\gamma, \tau)$ is the so-called automorphy factor~\cite{diamond2005first},
\begin{equation}
J_k(\gamma, \tau)\equiv(c\tau +d)^k, ~\quad~ \gamma=\begin{pmatrix}
a  ~& b  \\
c  ~& d
\end{pmatrix}\in \overline{\Gamma}\,.
\end{equation}
At the fixed point $\tau_0$, Eq.~\eqref{eq:MF_irr} gives us,
\begin{equation}
Y^{(k)}_{\mathbf{r}}(\tau_0)=Y^{(k)}_{\mathbf{r}}(\gamma_0\tau_0)=(c_0\tau_0+d_0)^k\rho_{\mathbf{r}}(\gamma_0)Y^{(k)}_{\mathbf{r}}(\tau_0)\,,
\end{equation}
which implies
\begin{equation}
\label{eq:MF_align}\rho_\mathbf{r}(\gamma_0)Y_\mathbf{r}(\tau_0)=J_k^{-1}(\gamma_0,\tau_0)Y_\mathbf{r}(\tau_0)\,,
\end{equation}
Hence the modular multiplets $Y_\mathbf{r}(\tau_0)$ at the fixed point $\tau_0$ is actually the eigenvector of the representation matrix $\rho_\mathbf{r}(\gamma_0)$ with eigenvalue $J_k^{-1}(\gamma_0,\tau_0)$. It is straightforward to obtain
\begin{eqnarray}
\nonumber&&S=\begin{pmatrix}
0 & 1 \\
-1  & 0
\end{pmatrix},~\quad~ J_{1}(S, \tau_S)=-i\,,\\
\nonumber&&ST=\begin{pmatrix}
0 & 1 \\
-1  & -1
\end{pmatrix},~\quad~ J_{1}(ST, \tau_{ST})=\omega^2\,, \\
\nonumber&&TS=\begin{pmatrix}
-1 & 1 \\
-1  & 0
\end{pmatrix},~\quad~ J_{1}(TS, \tau_{TS})=\omega^2\,, \\
\label{eq:Jgama0tau0}&&T=\begin{pmatrix}
1 & 1 \\
0 & 1
\end{pmatrix},~\quad~ J_{1}(T, \tau_T)=1\,.
\end{eqnarray}
Therefore the automorphy factor $J_k(\gamma_0,\tau_0)$ at the fixed points $\tau_S$, $\tau_{ST}$, $\tau_{TS}$ and $\tau_T$ must be a phase with unit absolute value. We find that the weight two modular forms at these fixed points take the following values
\begin{eqnarray}
\nonumber&&Y^{(2)}_{\mathbf{2}}(\tau_S)=Y_{S}\begin{pmatrix}
1\\-1
\end{pmatrix},~\quad~ Y^{(2)}_{\mathbf{3}}(\tau_S)=-\frac{\omega}{\sqrt{3}}Y_{S} \begin{pmatrix}
1 \\
1+\sqrt{6}\\
1-\sqrt{6}
\end{pmatrix}\,, \\
\nonumber&&Y^{(2)}_{\mathbf{2}}(\tau_{ST})=Y_{ST}\begin{pmatrix}
0\\
1
\end{pmatrix},~\quad~Y^{(2)}_{\mathbf{3}}(\tau_{ST})= \sqrt{3}\omega Y_{ST}\begin{pmatrix}
0\\
1\\
0
\end{pmatrix}\,, \\
\nonumber&&Y^{(2)}_{\mathbf{2}}(\tau_{TS})=Y_{TS}
\begin{pmatrix}
1\\
0
\end{pmatrix},~\quad~Y^{(2)}_{\mathbf{3}}(\tau_{TS})=-\frac{2\omega}{\sqrt{3}}Y_{TS} \begin{pmatrix}
1\\
1\\
-\frac{1}{2}
\end{pmatrix}\,,\\
&&Y^{(2)}_{\mathbf{2}}(\tau_{T})=Y_{T}\begin{pmatrix}
1\\
\omega
\end{pmatrix},~\quad~Y^{(2)}_{\mathbf{3}}(\tau_{T})=Y_{T} \begin{pmatrix}
1\\
\omega^{2}\\
\omega
\end{pmatrix}\,,
\end{eqnarray}
with $Y_{S}\simeq-1.045-0.603i$, $Y_{ST}\simeq1.793$, $Y_{TS}\simeq-0.896-1.553i$ and $Y_{T}\simeq-0.5-0.866i$. From Eq.~\eqref{eq:MF_align} we know that $Y^{(2)}_{\mathbf{3}}(\tau_S)$ is an eigenvector of the representation matrix $\rho_{\mathbf{3}}(S)$ with eigenvalue $-1$. However $\rho_{\mathbf{3}}(S)$ has two degenerate eigenvalues $-1$, consequently the alignment of $Y^{(2)}_{\mathbf{3}}(\tau_S)$ can not be uniquely fixed by Eq.~\eqref{eq:MF_align}. It is remarkable that modular symmetry helps to break the degeneracy in some sense and fix $Y^{(2)}_{\mathbf{3}}(\tau_S)$ along the direction $\left(1, 1+\sqrt{6}, 1-\sqrt{6}\right)^{T}$. The values of the modular forms of level 4 up to weight 6 at the fixed points $\tau_S$, $\tau_{ST}$, $\tau_{TS}$ and $\tau_T$ are summarized in table~\ref{Tab:MF@FP}. In the present work, we are mainly concerned with the triplet modular forms transforming as $\mathbf{3}$ or $\mathbf{3}'$.

\begin{table}[t!]
\centering
\resizebox{1.0\textwidth}{!}{
\begin{tabular}{|c|c|c|c|c|} \hline  \hline
& $\tau_{S}$ & $\tau_{ST}$ & $\tau_{TS}$ & $\tau_{T}$ \\ \hline
$Y^{(2)}_{\mathbf{2}}$ & $Y_{S}(1,-1)$ & $Y_{ST}(0,1)$ & $Y_{TS}(1,0)$ & $Y_{T}(1,\omega)$ \\ \hline
$Y^{(2)}_{\mathbf{3}}$ & $-\frac{\omega}{\sqrt{3}}Y_{S}(1, 1+\sqrt{6}, 1-\sqrt{6}) $ & $ \sqrt{3}\omega Y_{ST}(0, 1, 0)$ & $-\frac{2\omega}{\sqrt{3}}Y_{TS}(1,1,-\frac{1}{2}) $ & $ Y_{T}(1,\omega^{2},\omega)$ \\
\hline \hline
$Y^{(4)}_{\mathbf{1}}$ & $-2Y_{S}^{2}$ & $0$ & $0$ & $2\omega Y_{T}^{2}$ \\ \hline
$Y^{(4)}_{\mathbf{2}}$ & $Y_{S}^{2}(1,1)$ & $Y_{ST}^{2}(1,0)$ & $Y_{TS}^{2}(0,1)$  & $\omega^{2}Y_{T}^{2}(1,\omega)$\\ \hline
$Y^{(4)}_{\mathbf{3}}$ & $-2\sqrt{2}\omega Y_{S}^{2} (1,-\frac{1}{2},-\frac{1}{2})$ & $\sqrt{3}\omega Y_{ST}^{2}(0, 0, 1) $ & $-\frac{2\omega}{\sqrt{3}} Y_{TS}^{2} (1, -\frac{1}{2}, 1)$ & $2\omega^{2}Y_{T}^{2}(1,\omega^{2},\omega)$ \\
\hline
$Y^{(4)}_{\mathbf{3'}}$ & $-\frac{2\omega}{\sqrt{3}} Y_{S}^{2} (1, 1-\sqrt{\frac{3}{2}}, 1+\sqrt{\frac{3}{2}})$ & $-\sqrt{3}\omega  Y_{ST}^{2} (0,0,1)$ & $-\frac{2\omega}{\sqrt{3}} Y_{TS}^{2} (1, -\frac{1}{2}, 1) $ & $(0,0,0)$ \\ \hline \hline
$Y^{(6)}_{\mathbf{1}}$ & $0$ & $Y_{ST}^{3}$ & $Y_{TS}^{3}$ &  $2Y_{T}^{3}$\\ \hline
$Y^{(6)}_{\mathbf{1'}}$ & $-2Y_{S}^{3}$ & $Y_{ST}^{3}$ & $-Y_{TS}^{3}$ & $0$ \\ \hline
$Y^{(6)}_{\mathbf{2}}$ & $-2Y_{S}^{3}(1,-1)$ & $(0,0)$ & $(0,0)$ & $2\omega Y_{T}^{3}(1,\omega)$ \\ \hline
$Y^{(6)}_{\mathbf{3, I}}$ & $\frac{2\omega}{\sqrt{3}} Y_{S}^{3} (1, 1+\sqrt{6}, 1-\sqrt{6})$ & $(0,0,0)$ & $(0,0,0)$ & $2\omega Y_{T}^{3} (1,\omega^{2},\omega) $ \\ \hline
$Y^{(6)}_{\mathbf{3, II}}$ & $-\frac{2\omega}{\sqrt{3}} Y_{S}^{3} (1, 1-\sqrt{\frac{3}{2}}, 1+\sqrt{\frac{3}{2}})$ & $\sqrt{3}\omega Y_{ST}^{3} (1,0,0)$ & $\frac{\omega}{\sqrt{3}}Y_{TS}^{3}(1,-2,-2)$ & $2\omega Y_{T}^{3} (1,\omega^{2},\omega) $ \\ \hline
$Y^{(6)}_{\mathbf{3'}}$ & $-2 \sqrt{2}\omega Y_{S}^{3} (1,-\frac{1}{2}, -\frac{1}{2})$ & $\sqrt{3}\omega Y_{ST}^{3} (1,0,0)$ & $-\frac{\omega}{\sqrt{3}}Y_{TS}^{3}(1,-2,-2)$ & $(0,0,0)$ \\ \hline \hline
\end{tabular}}
\caption{\label{Tab:MF@FP}The values of the modular forms with weights $k=2,4,6$ and level 4 at the fixed points $\tau_{S}$, $\tau_{ST}$, $\tau_{TS}$ and $\tau_{T}$, where $Y_{S}\simeq-1.045-0.603i$, $Y_{ST}\simeq1.793$, $Y_{TS}\simeq-0.896-1.553i$ and $Y_{T}\simeq-0.5-0.866i$. }
\end{table}

\subsection{Fixed points in upper half complex plane }

Let us consider the general fixed points $\tau_{f}$ of the modular transformation $\gamma_f$ in the upper half complex plane. By definition, $\tau_f$ must be related to some modulus $\tau'\in\mathcal{F}$ by certain modular transformation $\gamma'$, i.e., $\tau_f=\gamma'\tau'$. Then the fixed point condition $\gamma_f\tau_f=\tau_f$ becomes
\begin{equation}
\gamma_f\gamma'\tau'=\gamma'\tau'\,.
\end{equation}
Multiplying $\gamma'^{-1}$ from left on both sides of the above equation, we can obtain
\begin{equation}
\gamma'^{-1}\gamma_f\gamma'\tau'=\tau'\,,
\end{equation}
which is exactly the fixed point condition Eq.~\eqref{eq:fp_equa} inside the fundamental domain. Hence we have
\begin{equation}
\begin{aligned}
\tau'&=\tau_0\in\left\{\tau_S, \tau_{ST}, \tau_{TS}, \tau_T\right\}\,,\\
\gamma'^{-1}\gamma_f\gamma'&=\gamma_0\in\left\{\pm S, \pm ST, \pm TS, \pm T\right\}\,.
\end{aligned}
\end{equation}
Hence the fixed point $\tau_f$ of the modular group and the corresponding stabilizer $\gamma_f$ are given by
\begin{equation}
\label{eq:fp_general}\tau_f=\gamma'\tau_0,~~\quad~~\gamma_f=\gamma'\gamma_0\gamma'^{-1},~~~~\gamma'\in\overline{\Gamma}\,,
\end{equation}
\begin{figure}[t!]
\begin{center}
\includegraphics[width=0.95\linewidth]{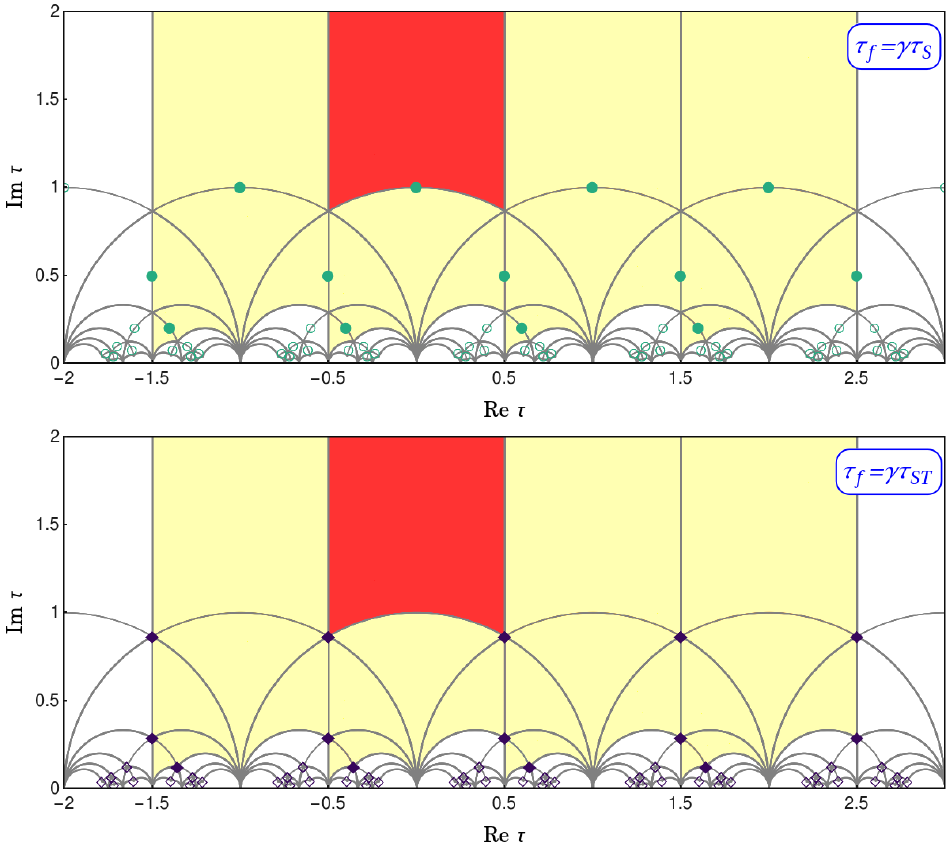}
\end{center}
\caption{\label{fig:fp}The fixed points of the modular group, it is impossible to display all of them because there are infinite fixed points. The red region and yellow region are the fundamental domains of $\overline{\Gamma}$ and $\overline{\Gamma}(4)$ respectively. The fixed points are displayed in solid (hollow) circles and diamonds in (outside) the fundamental domain of $\overline{\Gamma}(4)$. }
\end{figure}
where $\gamma'$ is an arbitrary modular symmetry element. It is straightforward to check $\gamma_f\tau_f=\tau_f$ which means that $\tau_f$ is really the fixed point of $\gamma_f$. Hence all fixed points are related to $\tau_S$, $\tau_{ST}$, $\tau_{TS}$, and  $\tau_T$ by modular transformations. For illustration, we display part of the fixed points of the modular group in figure~\ref{fig:fp}. Since there are infinity fixed points, it is impossible to show all the fixed points. From Eq.~\eqref{eq:fp_general} we see that only the modular symmetry transformation conjugate to $\gamma_0$ can have fixed point. In other words, $\gamma_f$ and $\gamma_0$ must belong to the same conjugacy class. The value of the modular form at the fixed point $\gamma_f$ is
\begin{equation}
\label{eq:Ytauf_bydef}Y_\mathbf{r}(\tau_f)= Y_\mathbf{r}(\gamma'\tau_0) = J_k(\gamma', \tau_0)\rho_\mathbf{r}(\gamma')Y_\mathbf{r}(\tau_0)\,,
\end{equation}
which implies that the direction of $Y_\mathbf{r}(\tau_f)$ is proportional to $\rho_\mathbf{r}(\gamma')Y_\mathbf{r}(\tau_0)$. The modular group $\overline{\Gamma}$ can be decomposed into disjoint union of right cosets of the principal congruence subgroup $\overline{\Gamma}(N)$ in $\overline{\Gamma}$,
\begin{equation}
\overline{\Gamma}=\bigcup_{i=1}^{|\Gamma_N|}g_i\overline{\Gamma}(N), ~\quad~ g_i \in \Gamma_N\,,
\end{equation}
where we taken into account the fact $\Gamma_N=\overline{\Gamma}/\overline{\Gamma}(N)$ such that the finite modular group $\Gamma_N$ can be regarded as the system of
right coset representatives of $\overline{\Gamma}(N)$ in $\overline{\Gamma}$. As a consequence, any element $\gamma\in\overline{\Gamma}$ can always be written as $\gamma=g_ih, h \in\overline{\Gamma}(N)$, thus $\rho_\mathbf{r}(\gamma)=\rho_\mathbf{r}(g_i)\rho_\mathbf{r}(h) = \rho_\mathbf{r}(g_i)$ because of the representation matrix $\rho_\mathbf{r}(h)=1$. Therefore the alignment $Y_\mathbf{r}(g_ih\tau_0)$ is proportional to $\rho_\mathbf{r}(g_i)Y_\mathbf{r}(\tau_0)$ and it is
independent of $h\in\overline{\Gamma}(N)$. The nonequivalent directions of the modular forms $Y_\mathbf{r}(\tau_f)$ at fixed points are actually generated by the finite modular group $\Gamma_N$. Although there are infinite numbers of fixed points $\tau_f$, it sufficient to only consider these fixed points $\tau_f$ which belong to the orbits $\Gamma_N\tau_S$, $\Gamma_N\tau_{ST}$, $\Gamma_N\tau_{TS}$ and $\Gamma_N\tau_{T}$. Moreover, for any fixed point $\tau_0=\tau_S$, $\tau_{ST}$, $\tau_{TS}$ or $\tau_{T}$ in the fundamental domain, the corresponding stabilizer subgroup is an abelian subgroup generated by $\gamma_0$ and it is denoted as $\texttt{Stab}_{\Gamma_N}(\tau_0)\equiv\langle\gamma_0\rangle$.
We can decompose the modular group $\Gamma_N$ into the right coset of the stabilizer subgroup $\texttt{Stab}_{\Gamma_N}(\tau_0)$: $\Gamma_N=\bigcup_i A_i~\texttt{Stab}_\Gamma(\tau_0)$ where $A_i$ is the right coset representative element. Thus the orbit $\Gamma_N\tau_0$ can be generally simplified into
\begin{equation}
\Gamma_N \tau_0 = \bigcup_i A_i~\texttt{Stab}_\Gamma(\tau_0) \tau_0 = \bigcup_i A_i \tau_0\,.
\end{equation}
Hence the number of the independent modulus in the orbit $\Gamma_N\tau_0$ is equal to $|\Gamma_N|/|\texttt{Stab}_\Gamma(\tau_0)|$ which is the number of distinct right cosets of $\texttt{Stab}_{\Gamma_N}(\tau_0)$. Here the notation $|G|$ denotes the order of a group $G$. Furthermore, after some algebra, we can show that the modular multiplet $Y_\mathbf{r}(\tau_f)$ has the following property
\begin{equation}
\label{eq:Ytauf}Y_\mathbf{r}(\tau_f)=J_k(\gamma_0,\tau_0)\rho_\mathbf{r}(\gamma_f)Y_\mathbf{r}(\tau_f)\,,
\end{equation}
which gives rise to
\begin{equation}
\rho_\mathbf{r}(\gamma_f)Y_\mathbf{r}(\tau_f) = J^{-1}_k(\gamma_0,\tau_0)Y_\mathbf{r}(\tau_f)\,.
\end{equation}
This means that the modular multiplet $Y_\mathbf{r}(\tau_f)$ at the fixed point $\tau_f$ is an eigenvector of the representation matrix $\rho_\mathbf{r}(\gamma_f)$ with the eigenvalue $J_k^{-1}(\gamma_0, \tau_0)$ given in Eq.~\eqref{eq:Jgama0tau0}. We give the nontrivial and nonequivalent fixed points and the corresponding alignments of the triplet modular forms of level 4 in table~\ref{vacuum_C24}. As we shall show in the present work, these alignments at the fixed points could give a rich phenomenology of neutrino mixing in the framework of tri-direct modular approach.

\begin{table}[t!]
\begin{center}
\resizebox{1.0\textwidth}{!}{
\begin{tabular}{|c|c|c|c|c|c|}\hline\hline
\multicolumn{6}{|c|}{The alignments of triplet modular forms $Y_{\mathbf{3}, \mathbf{3'}}(\gamma\tau_S)$ of level 4 up to weight 6}\\ \hline
$\gamma$ & $\gamma\tau_S$ & \multicolumn{2}{c|}{$Y^{(2)}_\mathbf{3}(\gamma\tau_S)$, $Y^{(6)}_\mathbf{3,I}(\gamma\tau_S)$} & $Y^{(4)}_\mathbf{3}(\gamma\tau_S)$, $Y^{(6)}_\mathbf{3'}(\gamma\tau_S)$ & $Y^{(4)}_\mathbf{3'}(\gamma\tau_S)$, $Y^{(6)}_\mathbf{3,II}(\gamma\tau_S)$ \\
\hline
$\{1, S\}$ & $i$ & \multicolumn{2}{c|}{$(1,1+\sqrt{6},1-\sqrt{6})$} & $(1,-\frac{1}{2},-\frac{1}{2})$ & $(1,1-\sqrt{\frac{3}{2}},1+\sqrt{\frac{3}{2}})$  \\
\hline
$\{T^2,T^2S\}$ & $2+i$ & \multicolumn{2}{c|}{$(1,\frac{1}{3}(-1+i\sqrt{2}),\frac{1}{3}(-1+i\sqrt{2}))$} & $(0,1,-1)$ & $(1,-\frac{i}{\sqrt{2}},-\frac{i}{\sqrt{2}})$  \\
\hline
$\{ST^2S,ST^2\}$ & $-\frac{2}{5}+\frac{i}{5}$ & \multicolumn{2}{c|}{$(1,-\frac{1}{3}(1+i\sqrt{2}),-\frac{1}{3}(1+i\sqrt{2}))$} & $(0,1,-1)$ & $(1,\frac{i}{\sqrt{2}},\frac{i}{\sqrt{2}})$  \\
\hline
$\{(ST^2)^2,T^2ST^2\}$ & $-\frac{8}{13}+\frac{i}{13}$ & \multicolumn{2}{c|}{$(1,1-\sqrt{6},1+\sqrt{6})$} & $(1,-\frac{1}{2},-\frac{1}{2})$ & $(1,1+\sqrt{\frac{3}{2}},1-\sqrt{\frac{3}{2}})$  \\ \hline
$\{ST,STS\}$ & $-\frac{1}{2}+\frac{i}{2}$ & \multicolumn{2}{c|}{$(1,\omega^2(1+\sqrt{6}),\omega(1-\sqrt{6}))$} & $(1,-\frac{\omega^2}{2},-\frac{\omega}{2})$ & $(1,\omega^2(1-\sqrt{\frac{3}{2}}),\omega(1+\sqrt{\frac{3}{2}}))$ \\
\hline
$\{TS,T\}$ & $1+i$ & \multicolumn{2}{c|}{$(1,-\frac{\omega}{3}(1+i\sqrt{2}),-\frac{\omega^2}{3}(1+i\sqrt{2}))$} & $(0,1,-\omega)$ & $(1,\frac{i\omega}{\sqrt{2}},\frac{i\omega^2}{\sqrt{2}})$  \\
\hline
$\{(ST)^2,T^3\}$ & $-1+i$ & \multicolumn{2}{c|}{$(1,\omega(1+\sqrt{6}),\omega(1-\sqrt{6}))$} & $(1,-\frac{\omega}{2},-\frac{\omega^2}{2})$ & $(1,\omega(1-\sqrt{\frac{3}{2})},\omega^2(1+\sqrt{\frac{3}{2}}))$ \\
\hline
$\{(TS)^2,TST\}$ & $\frac{1}{2}+\frac{i}{2}$ & \multicolumn{2}{c|}{$(1,\frac{\omega^2}{3}(-1+i\sqrt{2}),\frac{\omega}{3}(-1+i\sqrt{2}))$} & $(0,1,-\omega^2)$ & $(1,-\frac{i\omega^2}{\sqrt{2}},-\frac{i\omega}{\sqrt{2}})$  \\
\hline
$\{(T^2ST,TST^3\}$ & $\frac{3}{2}+\frac{i}{2}$ & \multicolumn{2}{c|}{$(1,\omega^2(1-\sqrt{6}),\omega(1+\sqrt{6}))$} & $(1,-\frac{\omega^2}{2},-\frac{\omega}{2})$ & $(1,\omega^2(1+\sqrt{\frac{3}{2}}),\omega(1-\sqrt{\frac{3}{2}}))$ \\
\hline
$\{(TST^2,TST^2S\}$ & $\frac{3}{5}+\frac{i}{5}$ & \multicolumn{2}{c|}{$(1,\omega(1-\sqrt{6}),\omega^2(1+\sqrt{6}))$} & $(1,-\frac{\omega}{2},-\frac{\omega^2}{2})$ & $(1,\omega(1+\sqrt{\frac{3}{2}}),\omega^2(1-\sqrt{\frac{3}{2}}))$  \\
\hline
$\{T^3ST^2,ST^2ST\}$ & $\frac{13}{5}+\frac{i}{5}$ & \multicolumn{2}{c|}{$(1,\frac{\omega}{3}(-1+i\sqrt{2}),\frac{\omega^2}{3}(-1+i\sqrt{2}))$} & $(0,1,-\omega)$ & $(1,-\frac{i\omega}{\sqrt{2}},-\frac{i\omega^2}{\sqrt{2}})$  \\
\hline
$\{T^2ST^3,T^3ST\}$ & $\frac{17}{10}+\frac{i}{10}$ & \multicolumn{2}{c|}{$(1,-\frac{\omega^2}{3}(1+i\sqrt{2}),-\frac{\omega}{3}(1+i\sqrt{2}))$} & $(0,1,-\omega^2)$ & $(1,\frac{i\omega^2}{\sqrt{2}},\frac{i\omega}{\sqrt{2}})$ \\
\hline\hline
\multicolumn{6}{|c|}{The alignments of triplet modular forms $Y_{\mathbf{3}, \mathbf{3'}}(\gamma\tau_{ST})$ of level 4 up to weight 6}\\ \hline
$\gamma$ & $\gamma\tau_{ST}$ & $Y^{(2)}_\mathbf{3}(\gamma\tau_{ST})$ & $Y^{(4)}_\mathbf{3}(\gamma\tau_{ST})$ , $Y^{(4)}_\mathbf{3'}(\gamma\tau_{ST})$ & $Y^{(6)}_{\mathbf{3},II}(\gamma\tau_{ST}), Y^{(6)}_\mathbf{3'}(\gamma\tau_{ST})$ & $Y^{(6)}_{\mathbf{3},I}(\gamma\tau_{ST})$ \\
\hline
$\{1, ST, (ST)^2\}$ & $\omega$ & $(0,1,0)$ & $(0,0,1)$ & $(1,0,0)$ & \multirow{8}*{$(0,0,0)$}\\
\cline{1-5}
$\{T^2,TS,T^2ST\}$ & $\omega+2$ & $(1,-\frac{\omega^2}{2},\omega)$ & $(1,\omega^2,-\frac{\omega}{2})$ & $(1,1+i\sqrt{3},1-i\sqrt{3})$& \\
\cline{1-5}
$\{ST^2S,(TS)^2,TST^2\}$ & $\frac{-5+i\sqrt{3}}{14}$ & $(1,-\frac{\omega}{2},\omega^2)$ & $(1,\omega,-\frac{\omega^2}{2})$ & $(1,1-i\sqrt{3},1+i\sqrt{3})$& \\
\cline{1-5}
$\{(ST^2)^2,T^3ST^2,T^2ST^3\}$ & $\frac{-9+i\sqrt{3}}{14}$ & $(1,-\frac{1}{2},1)$ & $(1,1,-\frac{1}{2})$ & $(1,-2,-2)$& \\
\cline{1-5}
$\{(S,T,TST\}$ & $-\omega^2$ & $(1,1,-\frac{1}{2})$ & $(1,-\frac{1}{2},1)$& $(1,-2,-2)$  &  \\
\cline{1-5}
$\{T^3ST,T^3,T^2S\}$ & $\omega+3$ & $(1,\omega,-\frac{\omega^2}{2})$ & $(1,-\frac{\omega}{2},\omega^2)$ & $(1,-2\omega,-2\omega^2)$ & \\
\cline{1-5}
$\{TST^3,T^2ST^2,TST^2S\}$ & $\frac{9+i\sqrt{3}}{14}$ & $(0,0,1)$ & $(0,1,0)$ & $(1,0,0)$  &  \\
\cline{1-5}
$\{ST^2ST,ST^2,STS\}$ & $\frac{-3+i\sqrt{3}}{6}$ & $(1,\omega^2,-\frac{\omega}{2})$ & $(1,-\frac{\omega^2}{2},\omega)$ &$(1,-2\omega^2,-2\omega)$ & \\
\hline\hline

\multicolumn{6}{|c|}{The alignments of triplet modular forms $Y_{\mathbf{3}, \mathbf{3'}}(\gamma\tau_{TS})$ of level 4 up to weight 6}\\ \hline

$\gamma$ & $\gamma\tau_{TS}$ & $Y^{(2)}_\mathbf{3}(\gamma\tau_{TS})$ & $Y^{(4)}_\mathbf{3}(\gamma\tau_{TS})$, $Y^{(4)}_\mathbf{3'}(\gamma\tau_{TS})$ & $Y^{(6)}_{\mathbf{3},II}(\gamma\tau_{TS})$, $Y^{(6)}_\mathbf{3'}(\gamma\tau_{TS})$  & $Y^{(6)}_{\mathbf{3},I
}(\gamma\tau_{TS})$ \\\hline
$\{1,TS,(TS)^{2}\}$ & $-\omega^2$& $(1,1,-\frac{1}{2})$ & $(1,-\frac{1}{2}, 1)$  & $(1,-2,-2)$ & \multirow{8}*{$(0,0,0)$}  \\ \cline{1-5}
$\{T^{2},(ST)^{2},T^{2}ST^{3}\}$ & $\frac{5+i\sqrt{3}}{2}$ & $(1,\omega,-\frac{1}{2}\omega^{2})$ & $(1,-\frac{1}{2}\omega,\omega^{2})$ & $(1,-2\omega,-2\omega^{2})$ & \\\cline{1-5}
$\{ST^{2}S,ST,T^{3}ST^{2}\}$ & $\frac{\sqrt{3}i-3}{6}$ & $(1,\omega^{2},-\frac{1}{2}\omega)$ & $(1,-\frac{1}{2}\omega^{2},\omega)$ & $(1,-2\omega^{2},-2\omega)$ & \\\cline{1-5}
$\{(ST^{2})^{2},T^{2}ST,TST^{2}\}$ & $\frac{\sqrt{3}i-23}{38}$ & $(0,0,1)$ & $(0,1,0)$ & $(1,0,0)$ & \\\cline{1-5}
$\{S,T^{3},STS\}$ & $\omega$ & $(0,1,0)$ & $(0,0,1)$ & $(1,0,0)$ & \\\cline{1-5}
$\{T^{3}ST,T^{2}ST^{2},ST^{2}ST\}$ & $3+\frac{(-1)^{5/6}}{\sqrt{3}}$ & $(1,-\frac{1}{2},1)$ & $(1,1,-\frac{1}{2})$ & $(1,-2,-2)$ & \\\cline{1-5}
$\{TST^{3},T,T^{2}S\}$ & $\frac{19+i\sqrt{3}}{26}$ & $(1,-\frac{1}{2}\omega^{2},\omega)$ & $(1,\omega^{2},-\frac{1}{2}\omega)$  & $(1,-2\omega^{2},2\omega)$ & \\\cline{1-5}
$\{TST^{2}S,ST^{2},TST\}$ & $\frac{3+i\sqrt{3}}{6}$ & $(1,-\frac{1}{2}\omega,\omega^{2})$ & $(1,\omega,-\frac{1}{2}\omega^{2})$ & $(1,-2\omega,-2\omega^{2})$ & \\ \hline\hline

\multicolumn{6}{|c|}{The alignments of triplet modular forms $Y_{\mathbf{3}, \mathbf{3'}}(\gamma\tau_{T})$ of level 4 up to weight 6}\\ \hline

$\gamma$ & $\gamma\tau_T$ & \multicolumn{3}{c|}{$Y^{(2)}_\mathbf{3}(\gamma\tau_T)$, $Y^{(4)}_\mathbf{3}(\gamma\tau_T)$, $Y^{(6)}_\mathbf{3,I}(\gamma\tau_T)$, $Y^{(6)}_\mathbf{3,II}(\gamma\tau_T)$} & $Y^{(4)}_\mathbf{3'}(\gamma\tau_T)$, $Y^{(6)}_\mathbf{3'}(\gamma\tau_T)$  \\\hline
  $\{1,T,T^{2},T^{3}\}$ & $i \infty$ & \multicolumn{3}{c|}{\multirow{2}*{$(1,\omega^{2},\omega)$}} & \multirow{6}*{$(0,0,0)$} \\\cline{1-2}
  $\{ST^{2}S,ST^{2}ST,(ST^{2})^{2},TST^{2}S\}$ & $-\frac{1}{2}$ & \multicolumn{3}{c|}{} &  \\\cline{1-5}

  $\{ST,(TS)^{2},S,ST^{2}\}$ & $0$ & \multicolumn{3}{c|}{\multirow{2}*{$(1, \omega,\omega^{2})$}} &  \\ \cline{1-2}
  $\{T^{2}ST,T^{2}ST^{3},T^{2}ST^{2},T^{2}S\}$ & $2$ & \multicolumn{3}{c|}{} & \\ \cline{1-5}

  $\{TS,TST^{2},TST^{3},TST\}$ & $1$ & \multicolumn{3}{c|}{\multirow{2}*{$(1,1,1)$}} & \\ \cline{1-2}
  $\{(ST)^{2},T^{3}ST^{2},T^{3}ST,STS\}$ & $-1$ & \multicolumn{3}{c|}{} & \\ \hline\hline

\end{tabular} }
\caption{\label{vacuum_C24}The alignments of the triplet modular forms $Y^{(2)}_{\mathbf{3}, \mathbf{3'}}(\tau_f)$, $Y^{(4)}_{\mathbf{3}, \mathbf{3'}}(\tau_f)$  and $Y^{(6)}_{\mathbf{3}, \mathbf{3'}}(\tau_f)$ of level 4 at the fixed point $\tau_f$. As shown in section~\ref{sec:fp&ResSym}, it is sufficient to only consider the fixed points $\tau_f=\gamma\tau_S$, $\gamma\tau_{ST}$, $\gamma\tau_{TS}$, $\gamma\tau_{T}$ with $\gamma \in S_4$. In the second column of the table, we have identified the modulus parameter $\tau$ with $T^4\tau=\tau+4$, the reason is $T^4=1$ in the $S_4$ group and $Y_{\mathbf{r}}(\tau)=Y_{\mathbf{r}}(\tau+4)$ for any modular multiplet $Y_{\mathbf{r}}$. }
\end{center}
\end{table}

\subsection{\label{subsec:Res_mod_symmetry}Residual modular symmetry and its implication }

We assume that neutrinos are Majorana particles in the following. The charged lepton and neutrino mass terms in modular invariant approach can be generally written as
\begin{equation}
\label{eq:Lag_mass}W_m=-(y_{e})_{ij}E^{c}_iY_{e}(\tau)L_jH_d-\frac{1}{2\Lambda}\left(y_{\nu}\right)_{ij}L_iL_jY_{\nu}(\tau)H_uH_u\,,
\end{equation}
where the two-component notation of the fermion fields is used. The fields $L_i$ and $E^c_i=e^{c}, \mu^{c}, \tau^{c}$ stand for the left-handed lepton doublets and the right-handed charged leptons respectively, $H_u$ and $H_d$ are Higgs doublets, and $Y_{e}(\tau)$ and $Y_{\nu}(\tau)$ are modular forms. The neutral components of the Higgs fields acquire vacuum expectation values $\langle H^0_{u,d}\rangle=v_{u, d}$ after the electroweak symmetry breaking, then the charged lepton and neutrino mass matrices can be read off as
\begin{equation}
m_e\left(\tau\right)=y_eY_e(\tau) v_d, ~\quad~ m_{\nu}\left(\tau\right)=y_{\nu}Y_{\nu}(\tau)\frac{v^2_u}{\Lambda}\,.
\end{equation}
Under a generic modular transformation $\gamma\in\overline{\Gamma}$, the lepton fields $L=\left(L_1, L_2, L_3\right)$ and $E^c=\left(e^c, \mu^c, \tau^c\right)^{T}$ transform as
\begin{eqnarray}
L\stackrel{\gamma}{\mapsto}(c\tau+d)^{-k_L}\rho_{L}(\gamma)L,~\quad~
E^c\stackrel{\gamma}{\mapsto}(c\tau+d)^{-k_E}\rho_{E}(\gamma)L\,,
\end{eqnarray}
where $-k_L$ and $-k_E$ are the modular weights of $L$ and $E^c$ respectively.  Modular invariance requires that the summation of the modular weights of each term in Eq.~\eqref{eq:Lag_mass} should be vanishing and the mass matrices $m_e\left(\tau\right)$ and $m_{\nu}\left(\tau\right)$ should transform under a modular transformation as
\begin{equation}
\begin{aligned}
&m_e\left(\tau\right)\stackrel{\gamma}{\mapsto}m_e\left(\gamma\tau\right)=(c\tau+d)^{k_L+k_E}\rho^*_{E}(\gamma)m_e\left(\tau\right)\rho^{\dagger}_L(\gamma)\,,\\
&m_\nu\left(\tau\right)\stackrel{\gamma}{\mapsto}m_\nu\left(\gamma\tau\right)=(c\tau+d)^{2k_L}\rho^*_{L}(\gamma)m_\nu\left(\tau\right)\rho^{\dagger}_L(\gamma)\,.
\end{aligned}
\end{equation}
In modular invariant models, the vacuum expectation value of the complex modulus $\langle\tau\rangle$ is the unique source of flavour symmetry breaking. If $\langle\tau\rangle$ is at some fixed point $\langle\tau\rangle=\tau_f$, a residual subgroup $\left\{\gamma^n_f: n\in\mathbb{Z}\right\}$ generated by the stabilizer $\gamma_f$ would be preserved. For example, the residual symmetries are $Z^{S}_2\equiv\left\{1, S\right\}$ and $Z^{ST}_3\equiv\left\{1, ST, (ST)^2\right\}$ respectively for $\langle\tau\rangle=\tau_S=i$ and $\langle\tau\rangle=\tau_{ST}=(-1+i\sqrt{3})/2$. In the following, we shall discuss the constraints on the lepton mass matrices $m_e\left(\tau\right)$ and $m_\nu\left(\tau\right)$ imposed by the residual modular symmetry. At the fixed point $\tau_f$ which fulfills $\tau_f=\gamma_f\tau_f$, we have
\begin{equation}
\label{eq:cons:ResModSym}
\begin{aligned}
&m_e(\tau_f)=m_e\left(\gamma_f\tau_f\right)=J_{k_L+k_E}(\gamma_f, \tau_f)\rho^*_{E}(\gamma_f)m_e\left(\tau_f\right)\rho^{\dagger}_L(\gamma_f)\,,\\
&m_\nu(\tau_f)=m_\nu\left(\gamma_f\tau_f\right)=J_{2k_L}(\gamma_f, \tau_f)
\rho^*_{L}(\gamma_f)m_\nu\left(\tau_f\right)\rho^{\dagger}_L(\gamma_f)\,,
\end{aligned}
\end{equation}
where $J_{k_L+k_E}(\gamma_f, \tau_f)=J_{k_L+k_E}(\gamma_0, \tau_0)$ and $J_{2k_L}(\gamma_f, \tau_f)=J_{2k_L}(\gamma_0, \tau_0)$ are certain phase factors~\footnote{Using the identities $J_{k}(\gamma_1\gamma_2, \tau)=J_{k}(\gamma_1, \gamma_2\tau)J_{k}(\gamma_2, \tau)$ and $J_{k}(\gamma^{-1}, \gamma\tau)=J^{-1}_k(\gamma, \tau)$, one can prove $J_{k}(\gamma_f, \tau_f)=J_{k}(\gamma'\gamma_0\gamma'^{-1}, \gamma'\tau_0)=J_{k}(\gamma_0, \tau_0)$}, as shown in Eq.~\eqref{eq:Jgama0tau0}. From Eq.~\eqref{eq:cons:ResModSym}, we know that the hermitian combination of the charged lepton mass matrix  $m^{\dagger}_e(\tau_f)m_e(\tau_f)$ should be subject to the following constraint,
\begin{equation}
\rho^{\dagger}_L(\gamma_f)m^{\dagger}_e(\tau_f)m_e(\tau_f)\rho_L(\gamma_f)=m^{\dagger}_e(\tau_f)m_e(\tau_f)\,,
\end{equation}
which implies $\rho_L(\gamma_f)$ and $m^{\dagger}_e(\tau_f)m_e(\tau_f)$ are commutable. The representation matrix $\rho_{L}(\gamma_f)$ can be diagonalized by a unitary matrix $U_{e}$,
\begin{equation}
\label{eq:Ue_Res}U^{\dagger}_e\rho_L(\gamma_f)U_{e}=\widehat{\rho}_L(\gamma_f)\,,
\end{equation}
where $\widehat{\rho}_L(\gamma_f)$ is a diagonal unitary matrix and the non-vanishing entries are eigenvalues of $\rho_L(\gamma_f)$. Then we can see
\begin{small}
\begin{equation}
U^{\dagger}_em^{\dagger}_e(\tau_f)m_e(\tau_f)U_{e}=U^{\dagger}_e\rho^{\dagger}_L(\gamma_f)m^{\dagger}_e(\tau_f)m_e(\tau_f)\rho_L(\gamma_f)U_e=
\widehat{\rho}^{\dagger}_L(\gamma_f)U^{\dagger}_em^{\dagger}_e(\tau_f)m_e(\tau_f)U_e\widehat{\rho}_L(\gamma_f)\,.
\end{equation}
\end{small}
If the diagonal entries of $\widehat{\rho}_L(\gamma_f)$ are non-degenerate, then $U^{\dagger}_em^{\dagger}_e(\tau_f)m_e(\tau_f)U_{e}$ is also diagonal since only a diagonal matrix is invariant when conjugated by a non-degenerate phase matrix. If $\widehat{\rho}_L(\gamma_f)$ has two degenerate diagonal entries, only one column of $U_{e}$ can be determined at the fixed point $\tau_f$. In the same fashion, we can show that the neutrino mass matrix $m_{\nu}(\tau_f)$ is also diagonalized by the unitary transformation $U_e$ in Eq.~\eqref{eq:Ue_Res}. As a result, the lepton mixing matrix would have six zeros for non-degenerate $\widehat{\rho}_L(\gamma_f)$ or four zeros for partially degenerate $\widehat{\rho}_L(\gamma_f)$~\cite{Novichkov:2018ovf,Novichkov:2018yse}, and this is not consistent with the experimental data. Therefore there are no phenomenologically viable models with one common fixed point $\tau_f$ in both neutrino and charged lepton sectors.

In order to accommodate the observed lepton mixing angles, it is assumed the complex modulus $\tau$ in the charged lepton and neutrino mass matrices takes two different values $\tau_{f,e}$ and $\tau_{f,\nu}$ which break the modular symmetry into the residual subgroups generated by $\gamma_{f,e}$ and $\gamma_{f,\nu}$ respectively in the charged lepton and neutrino sectors. In this approach, the trimaximal TM2 lepton mixing pattern can be obtained from the $A_4$ modular group~\cite{Novichkov:2018yse} and the models giving TM1 mixing are constructed with multiple modular $S_4$ groups~\cite{deMedeirosVarzielas:2019cyj,King:2019vhv}. Inspired by the tri-direct CP approach~\cite{Ding:2018fyz,Ding:2018tuj,Chen:2019oey}, in the following we shall present the tri-direct modular model which has three fixed moduli.

\section{\label{sec:tridirec_MM}New Predictive Examples of Lepton Mixing}

The tri-direct approach is based on the minimal seesaw model with two right-handed neutrinos which are denoted as $N^c_{\mathrm{atm}}$ and $N^c_{\mathrm{sol}}$~\cite{Ding:2018fyz,Ding:2018tuj,Chen:2019oey}. In this section we shall generalize the tri-direct approach to the modular invairance models, and the flavon fields would be replaced by modular forms. The superpotential for the charged lepton and neutrino masses in the tri-direct modular model is of the following form
\begin{eqnarray}
\nonumber W&=&-y_{e}E^{c}Y_{e}(\tau)LH_d-y_{\mathrm{atm}}LY_{\mathrm{atm}}(\tau)N^c_{\mathrm{atm}}H_u-y_{\mathrm{sol}}LY_{\mathrm{sol}}(\tau)N^c_{\mathrm{sol}}H_u \\
\label{eq:TD_MM_Lag}&&-\frac{1}{2}M_{\mathrm{atm}}N^c_{\mathrm{atm}}N^c_{\mathrm{atm}}
-\frac{1}{2}M_{\mathrm{sol}}N^{c}_{\mathrm{sol}}N^c_{\mathrm{sol}}\,,
\end{eqnarray}
where we don't specify whether the above operators arise from only one independent or several  independent modular invariant contractions. We assign the lepton doublets $L$ to triplet of the finite modular group $\Gamma_N$, while both right-handed neutrinos $N^c_{\mathrm{atm}}$ and $N^c_{\mathrm{sol}}$ transform as singlets of $\Gamma_N$. Then modular invariance requires $Y_{\mathrm{atm}}(\tau)$ and $Y_{\mathrm{sol}}(\tau)$ are triplet modular forms. For instance, in the case of $N=4$, we can assume $L\sim\mathbf{3}$, $N^c_{\mathrm{atm}}\sim\mathbf{1}$,  $N^c_{\mathrm{sol}}\sim\mathbf{1'}$, $Y_{\mathrm{atm}}(\tau)\sim\mathbf{3}$, $Y_{\mathrm{sol}}(\tau)\sim\mathbf{3'}$ or $L\sim\mathbf{3}$, $N^c_{\mathrm{atm}}\sim\mathbf{1'}$,  $N^c_{\mathrm{sol}}\sim\mathbf{1}$, $Y_{\mathrm{atm}}(\tau)\sim\mathbf{3'}$, $Y_{\mathrm{sol}}(\tau)\sim\mathbf{3}$. The cross term $N^{c}_{\mathrm{atm}}N^c_{\mathrm{sol}}$ could be forbidden by proper weight assignment of $N^c_{\mathrm{atm}}$ and $N^c_{\mathrm{sol}}$. Motivated by the idea of tri-direct CP approach~\cite{Ding:2018fyz,Ding:2018tuj,Chen:2019oey}, we shall consider the scenario of three moduli in the theory: $\tau_e$ and its vacuum expectation value (VEV) at certain fixed point $\tau_{f, e}$ break the modular symmetry in the charged lepton sector, $\tau_{\mathrm{atm}}$ and  $\tau_{\mathrm{sol}}$ are responsible for the breaking of modular symmetry in the atmospheric and solar neutrino sectors respectively.

Similar to section~\ref{subsec:Res_mod_symmetry}, the charged lepton mass matrix is determined to be $m_e(\tau_e)=y_eY_e(\tau_{e}) v_d$ and it is diagonalized by the unitary matrix $U_e$ fulfilling $U^{\dagger}_e\rho_L(\gamma_{f, e})U_{e}=\widehat{\rho}_L(\gamma_{f, e})$ at the fixed point $\langle\tau_e\rangle=\tau_{f,e}$. Notice that $U_{e}$ is determined up to permutations and phases of
its column vectors since the order of the charged lepton masses is undefined in our framework. Subsequently we can read out the neutrino Dirac mass matrix and the Majorana mass matrix of right-handed neutrinos,
\begin{equation}
m_{D}=\begin{pmatrix}
y_{\text{atm}}Y_{\text{atm}}\left(\tau_{\mathrm{atm}}\right)v_u,  & y_{\text{sol}}Y_{\text{sol}}\left(\tau_{\mathrm{sol}}\right)v_u
\end{pmatrix},\qquad m_{N}=\begin{pmatrix}
M_{\textrm{atm}} ~&~ 0 \\
0  ~& ~ M_{\textrm{sol}}
\end{pmatrix}\,,
\end{equation}
where the Clebsch-Gordan coefficients in both contractions $y_{\mathrm{atm}}LY_{\mathrm{atm}}N^c_{\mathrm{atm}}H_u$ and $y_{\mathrm{sol}}LY_{\mathrm{sol}}N^c_{\mathrm{sol}}H_u$ are omitted for notation simplicity. The effective light neutrino mass matrix is given by the seesaw formula $m_{\nu}=-m_Dm_N^{-1}m^T_D$ which implies
\begin{equation}
\label{eq:mnu_TDM}m_{\nu}=-\frac{y^2_{\mathrm{atm}}v^2_u}{M_{\mathrm{atm}}}Y_{\text{atm}}Y^{T}_{\text{atm}}-\frac{y^2_{\mathrm{sol}}v^2_u}{M_{\mathrm{sol}}}Y_{\text{sol}}Y^{T}_{\text{sol}}\,.
\end{equation}
We shall assume that the vacuum expectation values of $\tau_{\mathrm{atm}}$ and $\tau_{\mathrm{sol}}$ are certain fixed points $\langle\tau_{\mathrm{atm}}\rangle=\tau_{f, \mathrm{atm}}$ and $\langle\tau_{\mathrm{sol}}\rangle=\tau_{f, \mathrm{sol}}$ so that some residual modular groups generated by $\gamma_{f,\mathrm{atm}}$
and $\gamma_{f,\mathrm{sol}}$ are preserved. In order to show concrete examples and find new interesting models, we shall
perform a comprehensive study for the modular group $S_4$. Note that we leave the construction of tri-direct modular model in future work.

\subsection{Examples of $N=4$}

We have considered all possible fixed points of the three moduli $\tau_e$, $\tau_{\mathrm{atm}}$, $\tau_{\mathrm{sol}}$ and the corresponding alignments of triplets modular forms shown in table~\ref{vacuum_C24}. We find three possible cases which can accommodate the experimental data on neutrino masses and mixing angles. The assignments for the modular forms $Y_{\mathrm{atm}}$ and $Y_{\mathrm{sol}}$ and the vacuum expectation values of the moduli $\tau_{\mathrm{atm}}$ and $\tau_{\mathrm{sol}}$ are summarized in table~\ref{tab:assign_TDM}.

\begin{table}[t!]
\centering
\begin{tabular}{|c|c|c|c|c|c|}
  \hline
  \hline
 Case & $G_{e}$ & $Y_{\mathrm{atm}}$ & $Y_{\mathrm{sol}}$ & $\tau_{f, \mathrm{atm}}$ & $\tau_{f, \mathrm{sol}}$  \\ \hline

A & \multirow{2}{*}{$Z_{3}^{ST}$} &  $Y^{(2)}_{\mathbf{3}}$ & $ Y^{(6)}_{\mathbf{3'}}$ &  $\tau_S=i$  & $T^2\tau_{ST}=2+\omega$ \\ \cline{1-1} \cline{3-6}

 B  &   & $Y^{(4)}_{\mathbf{3}}, Y^{(6)}_{\mathbf{3'}}$   &  $Y^{(2)}_{\mathbf{3}}  $  &   $T^2\tau_S=2+i$  &  $\tau_S=i$  \\ \hline

C  &  $Z_{4}^{ST^{2}}$  &  $Y^{(2)}_{\mathbf{3}}$  & $Y^{(4)}_{\mathbf{3}}, Y^{(4)}_{\mathbf{3'}}$   &  $T^3\tau_S=-1+i$  &  $TST^{3}\tau_{ST}=\frac{9+i\sqrt{3}}{14}$  \\ \hline\hline

\end{tabular}
\caption{\label{tab:assign_TDM}The assignments for the modular forms $Y_{\mathrm{atm}}$ and $Y_{\mathrm{sol}}$ and the VEVs of $\tau_{\mathrm{atm}}$ and $\tau_{\mathrm{sol}}$ for the three phenomenologically viable cases of tri-direct modular approach with two right-handed neutrinos.}
\end{table}

\subsubsection{$G_{e}=Z_{3}^{ST}$}

\begin{figure}[t!]
\centering
\begin{tabular}{cc}
\includegraphics[width=0.48\linewidth]{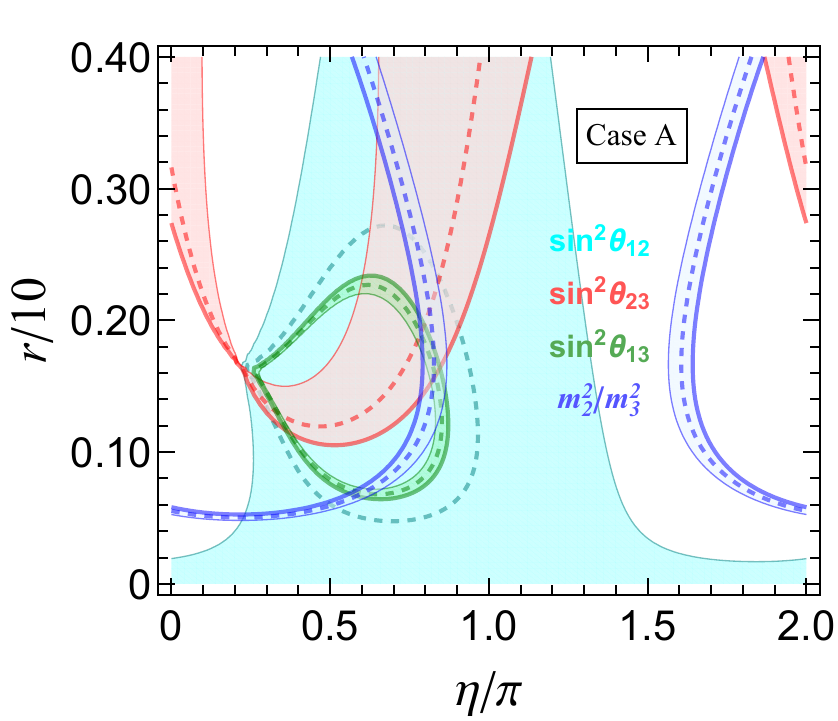}
\includegraphics[width=0.48\linewidth]{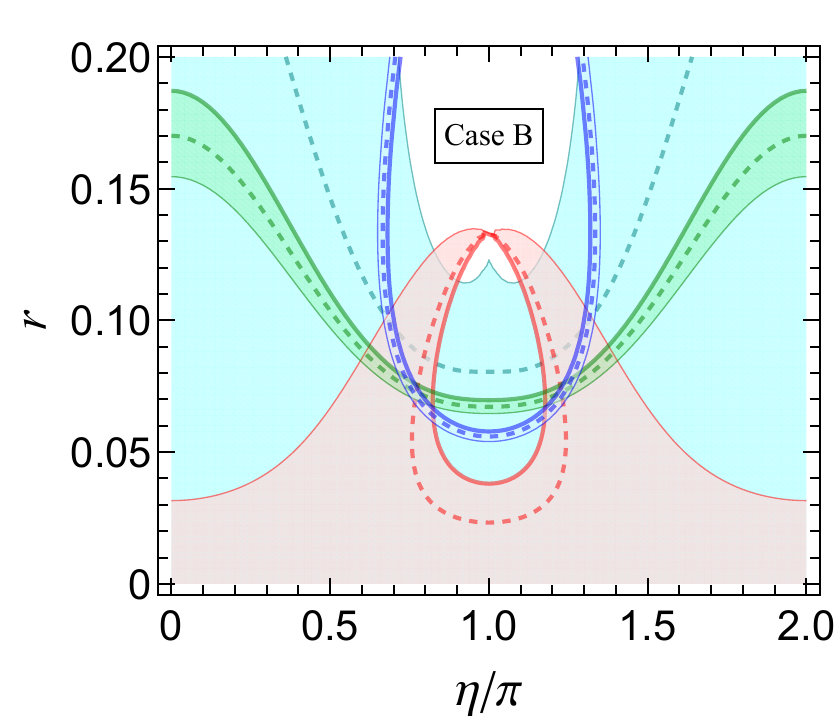}\\
\includegraphics[width=0.48\linewidth]{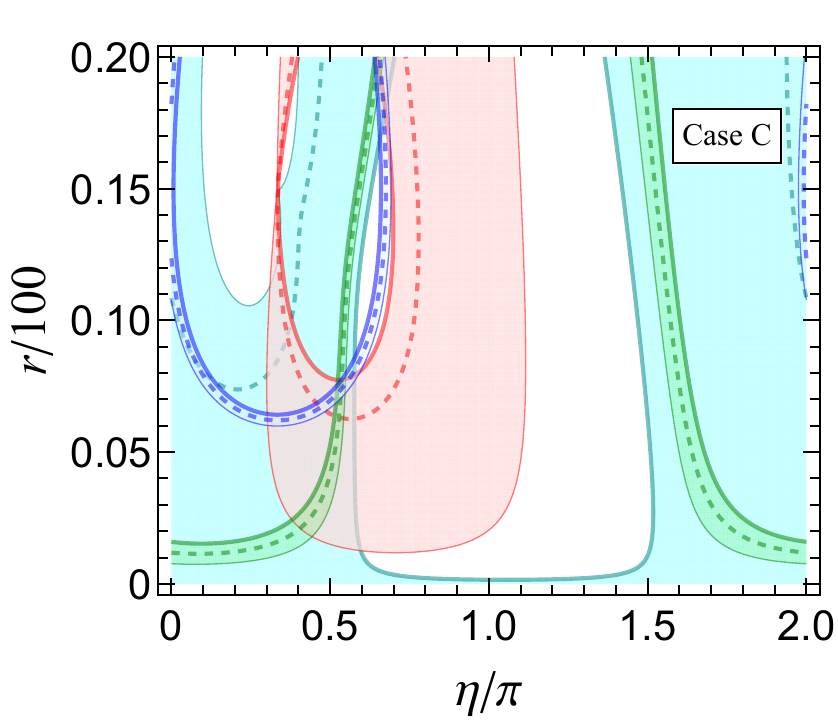}
\end{tabular}
\caption{\label{fig_contour:case_I} The contour plots of $\sin^2\theta_{12}$, $\sin^2\theta_{13}$, $\sin^{2}\theta_{23}$ and $m^2_{2}/m^2_{3}$ in the $\eta/\pi-r$ plane for cases $\mathbf{A}$, $\mathbf{B}$ and $\mathbf{C}$. The cyan, red, green and blue areas denote the $3\sigma$ regions of $\sin^{2}\theta_{23}$, $\sin^{2}\theta_{13}$ and $m_{2}^{2}/m_{3}^{2}$ respectively.
The solid lines denote the 3 sigma upper bounds, the thin lines denote the 3 sigma lower bounds
and the dashed lines refer to their best fit values, as adopted from NuFIT 4.1~\cite{Esteban:2018azc}.}
\end{figure}

If the VEV of the modulus $\tau_e$ is $\langle\tau_e\rangle=\tau_{ST}$, the modular symmetry would be broken down to the residual subgroup $G_{e}=Z_{3}^{ST}$ in the charged lepton sector. Since the representation matrix of the element $ST$ is diagonal, the charged lepton diagonalization matrix would be a unit matrix $U_{e}=\mathbb{1}_{3\times 3}$ up to column permutations, and the lepton flavour mixing completely originates from the neutrino sector. From table~\ref{vacuum_C24} we can read out the alignments of the triplet atmospheric and solar modular forms at the fixed points,
\begin{eqnarray}
&& \text{Case A:}~~Y_{\text{atm}}(\tau_{f, \mathrm{atm}}=i)\propto\begin{pmatrix}
1  \\
1+\sqrt{6}  \\
1-\sqrt{6} \\
\end{pmatrix},~~~Y_{\text{sol}}(\tau_{f, \mathrm{sol}}=2+\omega)\propto\begin{pmatrix}
1  \\
1+i\sqrt{3}  \\
1-i\sqrt{3} \\
\end{pmatrix},\\
&& \text{Case B:}~~ Y_{\text{atm}}(\tau_{f, \mathrm{atm}}=2+i)\propto\begin{pmatrix}
0  \\
1  \\
-1 \\
\end{pmatrix}, ~~~Y_{\text{sol}}(\tau_{f, \mathrm{sol}}=i)\propto\begin{pmatrix}
1  \\
1+\sqrt{6}  \\
1-\sqrt{6} \\
\end{pmatrix}\,.
\end{eqnarray}
Notice that the case B corresponds to the CSD($n$) model with $n=1+\sqrt{6}$, the two columns of the Dirac mass matrix are proportional to $\left(0, 1, -1\right)$ and $\left(1, n, 2-n\right)$ respectively in CSD($n$) model
\cite{King:2013iva,King:2013xba,King:2015dvf,Chen:2019oey}. The predictive Littlest seesaw model and its variant are the cases of $n=3$~\cite{King:2013iva,King:2015dvf,King:2016yvg,Ballett:2016yod,King:2018fqh},
$n=4$~\cite{King:2013xba,King:2013hoa,King:2014iia,Bjorkeroth:2014vha}
and $n=-1/2$~\cite{Chen:2019oey} respectively. It has been shown that the CSD($n$) model can be reproduced from the $S_4$ flavour symmetry in the tri-direct CP approach~\cite{Ding:2018fyz,Ding:2018tuj}, and the parameter $n$ is constrained to be a generic real parameter by the $S_4$ flavour symmetry and CP symmetry~\cite{Ding:2018fyz,Ding:2018tuj}. Here the modular symmetry can fix the alignment parameter $n$ to be $1+\sqrt{6}$. This is a remarkable advantage of modular symmetry with respect to discrete flavour symmetry. From Eq.~\eqref{eq:mnu_TDM}, we find the neutrino mass matrix is predicted to be
\begin{small}
\begin{eqnarray}
\nonumber&&\text{Case A}: m_{\nu}=m_a\left(
\begin{array}{ccc}
 1 & 1-\sqrt{6} & 1+\sqrt{6} \\
 1-\sqrt{6} & 7-2 \sqrt{6} & -5 \\
 1+\sqrt{6} & -5 & 7+2 \sqrt{6} \\
\end{array}
\right)+m_se^{i\eta}\left(
\begin{array}{ccc}
 1 & 1-i \sqrt{3} & 1+i \sqrt{3} \\
 1-i \sqrt{3} & -2-2 i \sqrt{3} & 4 \\
 1+i \sqrt{3} & 4 & -2+2 i \sqrt{3} \\
\end{array}
\right)\,,\\
&&\text{Case B}: m_{\nu}=m_a\left(
\begin{array}{ccc}
 0 & 0 & 0 \\
 0 & 1 & -1 \\
 0 & -1 & 1 \\
\end{array}
\right)+m_se^{i\eta}\left(
\begin{array}{ccc}
 1 & 1-\sqrt{6} & 1+\sqrt{6} \\
 1-\sqrt{6} & 7-2 \sqrt{6} & -5 \\
 1+\sqrt{6} & -5 & 7+2 \sqrt{6} \\
\end{array}
\right)\,.
\end{eqnarray}
\end{small}
It is notable that only three free parameters $m_a$, $m_s$ and $\eta$ are involved in the neutrino mass matrix. It is straightforward to check that the column vector $\left(2, -1, -1\right)^{T}$ is an eigenvector of $m_{\nu}$ with vanishing eigenvalue. Therefore the neutrino mass spectrum is normal ordering, and lightest neutrino is massless $m_1=0$, and the lepton mixing matrix is determined to be the TM1 pattern,
\begin{equation}
U_{PMNS}=\begin{pmatrix}
\sqrt{\frac{2}{3}}  &  -  &  -  \\
-\frac{1}{\sqrt{6}}  &  -  &  -   \\
-\frac{1}{\sqrt{6}}  &  -  &  -
\end{pmatrix}\,.
\end{equation}
Using the general formulae for lepton mixing angles, CP violating phases and neutrino masses in the tri-direct model given in~\cite{Ding:2018fyz,Ding:2018tuj}, we find the experimental data can be accommodated very well for certain values of $m_a$, $m_s$ and $\eta$. We present the results of $\chi^2$ analysis for the different cases in table~\ref{tab:bf_Z3}. The contour plots of $\sin^2\theta_{13}$, $\sin^2\theta_{12}$, $\sin^2\theta_{23}$ and $m^2_2/m^2_3$ in the plane $r$ versus $\eta$ are shown in figure~\ref{fig_contour:case_I} where $r=m_s/m_a$.

\begin{table}[t!]
\centering
\resizebox{1.00\textwidth}{!}{
\begin{tabular}{|c|c| c| c| c | c| c| c| c| c| c|c |c|c|c|}
  \hline \hline
  Case   & $\eta/\pi$  & $m_{a}(\text{meV})$ & $m_s/m_a$    & $\chi^2_{\text{min}}$ &  $\sin^2\theta_{13}$  &$\sin^2\theta_{12}$  & $\sin^2\theta_{23}$  & $\delta_{CP}/\pi$ &  $\beta/\pi$ & $m_2(\text{meV})$ & $m_3(\text{meV})$ & $m_{ee}(\text{meV})$ \\
  \hline
  A &  0.818 & 2.166 & 1.722 & 3.804 & 0.0227 & 0.318 & 0.591 & $-0.369$ & $-0.564$ & 8.697 & 50.166 & 2.241\\
  \hline
  B & 0.758 & 31.355 & 0.074 & 0.609 & 0.0223 & 0.318 & 0.553 & $-0.426$ & $-0.473$ & 8.610 & 50.264 & 2.300\\
  \hline

C & 0.533 & 4.352 & 7.334 & 11.188 & 0.0226 & 0.346 & 0.596 & $-0.365$ & $-0.608$ & 8.565 & 50.312 & 2.543 \\
\hline \hline
\end{tabular}}
\caption{\label{tab:bf_Z3} Results of the $\chi^2$ analysis for cases A, B and C. Here $\chi^2_{\text{min}}$ is the global minimum of the $\chi^2$ function. The parameter $\beta$ denotes the Majorana CP violation phase, the predictions for the effective Majorana mass $m_{ee}$ in neutrinoless double decay are listed in the last column. Note that the lightest neutrino is massless $m_1=0$ for each case. }
\end{table}

\subsubsection{$G_{e}=Z_{4}^{ST^{2}}$}

The residual symmetry of the charged lepton sector would be $G_e=Z_{4}^{ST^{2}}$ for the fixed point $\tau_{f, e}=-1$. In the case, the unitary transformation $U_{e}$ is of the following form,
\begin{equation}
U_{e}=\frac{1}{6}\left( \begin{array}{ccc}
2\sqrt{3} ~& 2\sqrt{3}  ~& -2\sqrt{3}  \\
 -3-\sqrt{3} ~& 2\sqrt{3}  ~& -3+\sqrt{3}  \\
3-\sqrt{3}  ~& 2\sqrt{3}  ~& 3+\sqrt{3}
\end{array} \right)\,.
\end{equation}
There are only one independent case which can be compatible with experimental data. The triplet modular forms $Y_{\mathrm{atm}}$ and $Y_{\mathrm{sol}}$ are aligned along the directions
\begin{eqnarray}
&& \hskip-0.4in \text{Case C:}~Y_{\text{atm}}(\tau_{f, \mathrm{atm}}=-1+i)\propto\begin{pmatrix}
1  \\
(1+\sqrt{6})\omega  \\
(1-\sqrt{6})\omega^{2} \\
\end{pmatrix},~~ Y_{\text{sol}}(\tau_{f, \mathrm{sol}}=((9+i\sqrt{3})/14)\propto\begin{pmatrix}
0  \\
1  \\
0 \\
\end{pmatrix}.
\end{eqnarray}
The most general form of the neutrino mass matrix reads as
\begin{equation}
m_{\nu}=m_a\begin{pmatrix}
1  & (1-\sqrt{6})\omega^{2}  & (1+\sqrt{6})\omega  \\
(1-\sqrt{6})\omega^{2}  & (7-2\sqrt{6})\omega  &  -5 \\
(1+\sqrt{6})\omega  &  -5 & (7+2\sqrt{6})\omega^2
\end{pmatrix}+m_se^{i\eta}\begin{pmatrix}
0 & 0 & 0  \\
0 & 0 & 0 \\
0 & 0 & 1
\end{pmatrix}\,.
\end{equation}
The lepton mixing matrix turns out to be of the form
\begin{equation}
U_{PMNS}=\begin{pmatrix}
\frac{1}{2\sqrt{10}}\sqrt{10+(1+\sqrt{2})(4+\sqrt{6})}  &  -  & - \\
\frac{1}{2\sqrt{5}}\sqrt{6-\sqrt{6}}  &  -  & - \\
\frac{1}{2\sqrt{10}}\sqrt{10+(1-\sqrt{2})(4+\sqrt{6})}  &  -  & -
\end{pmatrix}\simeq\begin{pmatrix}
0.800  &  -  & - \\
0.421  &  -  & - \\
0.428  &  -  & -
\end{pmatrix}\,.
\end{equation}
Notice that the first column of the mixing matrix looks much more complex although the experimental data can be accommodated. The predictions for lepton mixing parameters and neutrino masses are listed in table~\ref{tab:bf_Z3}.

For the cases A, B and C, both $Y_{\mathrm{atm}}$ and $Y_{\mathrm{sol}}$ are triplet modular forms with weights 2 or 4, and their values at fixed points are uniquely fixed, as shown in table~\ref{vacuum_C24}. However, at weight 6 there are two linearly independent modular forms $Y^{(6)}_{\mathbf{3}, I}$ and $Y^{(6)}_{\mathbf{3}, II}$ transforming as $\mathbf{3}$ under $S_4$. As a consequence, if either $Y_{\mathrm{atm}}$ or $Y_{\mathrm{sol}}$ is a weight 6 modular form in the representation $\mathbf{3}$, it alignment would not be fixed uniquely by the residual modular symmetry. For instance, we assign $G_{e}=Z^{ST}_3$ and
\begin{equation}
Y_{\mathrm{atm}}\sim Y^{(4)}_{\mathbf{3}}~\text{or}~Y^{(6)}_{\mathbf{3'}},\quad Y_{\mathrm{sol}}\sim Y^{(6)}_{\mathbf{3}},~~
\tau_{f, \mathrm{atm}}=T^2\tau_S=2+i,\quad \tau_{f, \mathrm{sol}}=ST\tau_S=-\frac{1}{2}+\frac{i}{2}\,,
\end{equation}
which implies
\begin{equation}
\hskip-0.1inY_{\mathrm{atm}}\left(\tau_{f, \mathrm{atm}}\right)\propto\begin{pmatrix}
0 \\
1 \\
-1
\end{pmatrix},~Y^{(6)}_{\mathbf{3}, I}\left(\tau_{f, \mathrm{sol}}\right)\propto\begin{pmatrix}
1  \\
(1+\sqrt{6})\omega^{2}  \\
(1-\sqrt{6})\omega \\
\end{pmatrix},~Y^{(6)}_{\mathbf{3}, II}\left(\tau_{f, \mathrm{sol}}\right)\propto\begin{pmatrix}
1  \\
(1-\sqrt{\frac{3}{2}})\omega^{2}  \\
(1+\sqrt{\frac{3}{2}})\omega \\
\end{pmatrix}\,.
\end{equation}
The solar modular form $Y_{\mathrm{sol}}$ is generally a linear combination of $Y^{(6)}_{\mathbf{3}, I}$ and $Y^{(6)}_{\mathbf{3}, II}$, and it can be written as
\begin{equation}
Y_{\mathrm{sol}}\left(\tau_{f, \mathrm{sol}}\right)=r_1e^{i\eta_1}Y^{(6)}_{\mathbf{3}, I}\left(\tau_{f, \mathrm{sol}}\right)+r_2e^{i\eta_2}Y^{(6)}_{\mathbf{3}, II}\left(\tau_{f, \mathrm{sol}}\right)\,,
\end{equation}
where $r_{1,2}$ and $\eta_{1,2}$ are real parameters. Then the light neutrino mass matrix can be parameterized as
\begin{equation}
\label{eq:mnu_3deg}m_{\nu}=m_{a}\left[Y_{\mathrm{atm}}\left(\tau_{f, \mathrm{atm}}\right)Y^{T}_{\mathrm{atm}}\left(\tau_{f, \mathrm{atm}}\right)+Y_{\mathrm{sol}}\left(\tau_{f, \mathrm{sol}}\right)Y^{T}_{\mathrm{sol}}\left(\tau_{f, \mathrm{sol}}\right)\right]\,.
\end{equation}
In this scenario, the neutrino mass matrix depends on five free parameters $m_a$, $r_1$, $r_2$, $\eta_1$ and $\eta_2$. The experimental data can be reproduced very well, e.g.
\begin{equation}
m_a=20.362\,\mathrm{meV},~~r_1=0.826,~~\eta_1=-0.588\pi, ~~r_2= 0.427,~~\eta_2=0.062\pi\,,
\end{equation}
and the best fit values of the neutrino mixing angles and mass squared differences can be obtained,
\begin{equation}
\begin{aligned}
&\sin^2\theta_{13}=0.0224,~~\sin^2\theta_{12}=0.310,~~\sin^2\theta_{23}=0.563,~~\delta_{CP}=-0.413\pi,\\
&\beta=0.556\pi,~~m_1=0\,\mathrm{meV},~~m_2=8.597\,\mathrm{meV},~~m_3=50.279\,\mathrm{meV}\,.
\end{aligned}
\end{equation}
Choosing other possible values of the fixed points $\tau_{f, e}$, $\tau_{f, \mathrm{atm}}$ and $\tau_{f, \mathrm{sol}}$, we can find similar models compatible with data. We shall not list all the possibilities since there are many viable cases.

\subsection{Tri-direct modular model with three right-handed neutrinos }

The above tri-direct modular approach can be straightforwardly extended to the conventional seesaw model with three right-handed neutrinos denote as $N^c_{\mathrm{atm}}$, $N^c_{\mathrm{sol}}$ and $N^c_{\mathrm{dec}}$. Then the superpotential for the charged lepton and neutrino masses is
\begin{eqnarray}
\nonumber W&=&-y_{e}E^{c}Y_{e}(\tau)LH_d-y_{\mathrm{atm}}LY_{\mathrm{atm}}(\tau)N^c_{\mathrm{atm}}H_u-y_{\mathrm{sol}}LY_{\mathrm{sol}}(\tau)N^c_{\mathrm{sol}}H_u
-y_{\mathrm{dec}}LY_{\mathrm{dec}}(\tau)N^c_{\mathrm{dec}}H_u\\
\label{eq:TD_MM3_Lag}&&-\frac{1}{2}M_{\mathrm{atm}}N^c_{\mathrm{atm}}N^c_{\mathrm{atm}}
-\frac{1}{2}M_{\mathrm{sol}}N^{c}_{\mathrm{sol}}N^c_{\mathrm{sol}}-\frac{1}{2}M_{\mathrm{dec}}N^{c}_{\mathrm{dec}}N^c_{\mathrm{dec}}\,.
\end{eqnarray}
In this scenario, the theory has four moduli: $\tau_e$ in the charged lepton sector, $\tau_{\mathrm{atm}}$, $\tau_{\mathrm{sol}}$ and $\tau_{\mathrm{dec}}$ associated with the atmospheric, solar and ``decoupled'' right-handed neutrinos sectors respectively. Analogous to the two right-handed neutrinos case discussed above, we assume the modular symmetry is broken to different residual subgroups through the VEVs of $\tau_e$, $\tau_{\mathrm{atm}}$, $\tau_{\mathrm{sol}}$ and $\tau_{\mathrm{dec}}$. Many viable model can be found in this scenario. For illustration, we shall present one example in the following, and we assume the residual modular symmetry $G_e=Z^{ST}_3$ in the charged lepton sector and
\begin{equation}
\begin{aligned}
&Y_{\mathrm{atm}}\sim Y^{(4)}_{\mathbf{3}}, Y^{(6)}_{\mathbf{3}'},~~~~Y_{\mathrm{sol}}\sim Y^{(2)}_{\mathbf{3}},~~~~Y_{\mathrm{dec}}\sim Y^{(4)}_{\mathbf{3}'},\\
&\tau_{f, \mathrm{atm}}=T^2\tau_S=2+i,~~\tau_{f, \mathrm{sol}}=\tau_S=i,~~\tau_{f, \mathrm{dec}}=T\tau_S=1+i\,.
\end{aligned}
\end{equation}
From table~\ref{vacuum_C24} we can read off the alignments of the modular forms as follow
\begin{equation}
Y_{\mathrm{atm}}(\tau_{f, \mathrm{atm}})\propto\begin{pmatrix}
0  \\
1  \\
-1 \\
\end{pmatrix},~~~Y_{\mathrm{sol}}(\tau_{f, \mathrm{sol}})\propto\begin{pmatrix}
1  \\
1+\sqrt{6}  \\
1-\sqrt{6} \\
\end{pmatrix},~~~Y_{\mathrm{dec}}(\tau_{f, \mathrm{dec}})\propto\begin{pmatrix}
1  \\
\frac{i}{\sqrt{2}}\omega  \\
\frac{i}{\sqrt{2}}\omega^{2} \\
\end{pmatrix}\,.
\end{equation}
Consequently the neutrino mass matrix is given by
\begin{eqnarray}
\nonumber m_{\nu}&=&m_a\left[\left(
\begin{array}{ccc}
 0 & 0 & 0 \\
 0 & 1 & -1 \\
 0 & -1 & 1 \\
\end{array}
\right)+r_1e^{i\eta_1}\left(
\begin{array}{ccc}
 1 & 1-\sqrt{6} & 1+\sqrt{6} \\
 1-\sqrt{6} & 7-2 \sqrt{6} & -5 \\
 1+\sqrt{6} & -5 & 7+2 \sqrt{6} \\
\end{array}
\right)\right.\\
&&~~\left.+r_2e^{i\eta_2}\frac{1}{2}\begin{pmatrix}
2  &  \sqrt{2}i\omega^2  &  \sqrt{2}i\omega \\
\sqrt{2}i\omega^2  & -\omega  &  -1 \\
 \sqrt{2}i\omega   &  -1   &  -\omega^2
\end{pmatrix}\right]\,,
\end{eqnarray}
where $r_{1,2}$ and $\eta_{1,2}$ are real free parameters. Excellent agreement with experimental data can be achieved in this case, and a numerical benchmark is
\begin{eqnarray}
\nonumber &&\hskip-0.15in  m_{a}=16.629\,\text{meV}, \quad  r_{1}=0.150, \quad r_{2}=0.877, \quad\eta_{1}=1.313\pi,\quad \eta_2=0.787 \pi\,,\\
\nonumber && \hskip-0.15in  \sin^2\theta_{13}=0.0224,\quad  \sin^2\theta_{12}=0.310, \quad \sin^2\theta_{23}=0.563\,,\\
\nonumber && \hskip-0.15in   \delta_{CP}=-0.635\pi, \quad \alpha_{21}=0.813\pi,\quad \alpha_{31}=-0.547\pi \,, \\
&& \hskip-0.15in     m_1= 3.830\,\text{meV}, \quad m_2=9.411\,\text{meV},  \quad m_3=50.425\,\text{meV}\,.
\end{eqnarray}

\section{\label{sec:Tri-direct_MM_A4}Fixed points and tri-direct modular model for $N=3$}

The modular group $\Gamma(3)$ has been extensively studied in the literature~\cite{Feruglio:2017spp,Criado:2018thu,Kobayashi:2018vbk,Kobayashi:2018scp,Okada:2018yrn,Kobayashi:2018wkl,Novichkov:2018yse,Nomura:2019yft}. The finite modular group $\Gamma_3$ is isomorphic to $A_4$. In the present work we shall adopt the same convention as~\cite{Feruglio:2017spp,Ding:2019zxk}. In the triplet representation $\mathbf{3}$, the $A_4$ generators $S$ and $T$ are represented by
\begin{equation}
S=\frac{1}{3}\left(\begin{array}{ccc}
-1& 2  & 2  \\
2  & -1  & 2 \\
2 & 2 & -1
\end{array}\right), ~~~~
T=\left(\begin{array}{ccc}
1 ~&~ 0 ~&~ 0 \\
0 ~&~ \omega^{2} ~&~ 0 \\
0 ~&~ 0 ~&~ \omega
\end{array}\right) \,.
\end{equation}
The weight 2 modular forms $Y^{(2)}_{\mathbf{3}}=\left(Y_1, Y_2, Y_3\right)^{T}$ transform as a triplet of $A_4$ and it can be constructed in terms of $\eta(\tau)$ as follow~\cite{Feruglio:2017spp},
\begin{eqnarray}
Y_1(\tau) &=& \frac{i}{2\pi}\left[ \frac{\eta'(\tau/3)}{\eta(\tau/3)}  +\frac{\eta'((\tau +1)/3)}{\eta((\tau+1)/3)}
+\frac{\eta'((\tau +2)/3)}{\eta((\tau+2)/3)} - \frac{27\eta'(3\tau)}{\eta(3\tau)}  \right], \nonumber \\
Y_2(\tau) &=& \frac{-i}{\pi}\left[ \frac{\eta'(\tau/3)}{\eta(\tau/3)}  +\omega^2\frac{\eta'((\tau +1)/3)}{\eta((\tau+1)/3)}
+\omega \frac{\eta'((\tau +2)/3)}{\eta((\tau+2)/3)}  \right] ,\nonumber \\
Y_3(\tau) &=& \frac{-i}{\pi}\left[ \frac{\eta'(\tau/3)}{\eta(\tau/3)}  +\omega\frac{\eta'((\tau +1)/3)}{\eta((\tau+1)/3)}
+\omega^2 \frac{\eta'((\tau +2)/3)}{\eta((\tau+2)/3)} \right]\,.
\end{eqnarray}
There are five weight 4 modular forms given by
\begin{equation}
\begin{aligned}
&Y^{(4)}_{\mathbf{1}}=Y_1^2+2 Y_2 Y_3\sim\mathbf{1},\\
&Y^{(4)}_{\mathbf{1}'}=Y_3^2+2 Y_1 Y_2\sim\mathbf{1}'\,,\\
&Y^{(4)}_{\mathbf{3}}=\left(\begin{array}{c}
Y_1^2-Y_2 Y_3\\
Y_3^2-Y_1 Y_2\\
Y_2^2-Y_1 Y_3
\end{array}
\right)\sim\mathbf{3}\,.
\end{aligned}
\end{equation}
The weight 6 modular forms can be decomposed as $\mathbf{1}\oplus\mathbf{3}\oplus\mathbf{3}$ under $A_4$~\cite{Feruglio:2017spp},
\begin{equation}
\begin{aligned}
&Y^{(6)}_{\mathbf{1}}=Y_1^3+Y_2^3+Y_3^3-3 Y_1 Y_2 Y_3\sim\mathbf{1}\,,\\
&Y^{(6)}_{\mathbf{3}, I}=\left(\begin{array}{c}
Y_1^3+2 Y_1Y_2Y_3\\
Y_1^2Y_2+2Y_2^2Y_3 \\
Y_1^2Y_3+2Y_3^2Y_2
\end{array}\right)\,,\\
&Y^{(6)}_{\mathbf{3}, II}=\left(\begin{array}{c}
Y_3^3+2 Y_1Y_2Y_3\\
Y_3^2Y_1+2Y_1^2Y_2\\
Y_3^2Y_2+2Y_2^2Y_1
\end{array}\right)\,.
\end{aligned}
\end{equation}
As shown in section~\ref{sec:fp&ResSym}, the modular group has infinite nontrivial fixed points while the nonequivalent alignments of the modular forms at fixed points are finite. It is sufficient to only consider the fixed points $\Gamma_N\tau_S$, $\Gamma_N\tau_{ST}$, $\Gamma_N\tau_{TS}$ and $\Gamma_N\tau_{T}$. We report the nonequivalent fixed points and the alignments of the triplet modular forms for $N=3$ in table~\ref{tab:FP&vacuum_A4}.

\begin{table}[t!]
\begin{center}
\resizebox{0.88\textwidth}{!}{
\begin{tabular}{|c|c|c|c|c|c|}\hline\hline
\multicolumn{6}{|c|}{The alignments of triplet modular forms $Y_{\mathbf{3}, \mathbf{3'}}(\gamma\tau_S)$ of level 3 up to weight 6}\\ \hline
$\gamma$ & $\gamma\tau_S$  & $Y^{(2)}_\mathbf{3}(\gamma\tau_S)$, $Y^{(6)}_{\mathbf{3}, I}(\gamma\tau_S)$ & $Y^{(4)}_\mathbf{3}(\gamma\tau_S)$  & \multicolumn{2}{c|}{$Y^{(6)}_{\mathbf{3}, II}(\gamma\tau_S)$} \\\hline
$\{1,S\}$ & $i$ & $(1,1-\sqrt{3},\sqrt{3}-2)$ & $(1,1,1)$ & \multicolumn{2}{c|}{$(1,-2-\sqrt{3},1+\sqrt{3})$} \\\hline
$\{T,TS\}$ & $1+i$ & $(1,(1-\sqrt{3})\omega,(\sqrt{3}-2)\omega^{2})$ & $(1,\omega,\omega^{2})$ & \multicolumn{2}{c|}{$(1,(-2-\sqrt{3})\omega,(1+\sqrt{3})\omega^{2})$} \\\hline
$\{ST,STS\}$ & $\frac{-1+i}{2}$ & $(1,(1+\sqrt{3})\omega,(-2-\sqrt{3})\omega^{2})$ & $(1,\omega,\omega^{2})$ & \multicolumn{2}{c|}{$(1,(\sqrt{3}-2)\omega,(1-\sqrt{3})\omega^{2})$} \\
    \hline
  $\{T^{2},T^{2}S\}$ & $2+i$ & $(1,(1-\sqrt{3})\omega^{2},(-2+\sqrt{3})\omega)$ & $(1,\omega^{2},\omega)$ & \multicolumn{2}{c|}{$(1,(-2-\sqrt{3})\omega^{2},(1+\sqrt{3})\omega)$} \\
    \hline
  $\{ST^{2},ST^{2}S\}$ & $\frac{-2+i}{5}$ & $(1,(1+\sqrt{3})\omega^{2},(-2-\sqrt{3})\omega)$ & $(1,\omega^{2},\omega)$ & \multicolumn{2}{c|}{$(1,(\sqrt{3}-2)\omega^{2},(1-\sqrt{3})\omega)$} \\
    \hline
  $\{T^{2}ST,TST^{2}\}$ & $\frac{3+i}{2}$ & $(1,1+\sqrt{3},-2-\sqrt{3})$ & $(1,1,1)$ & \multicolumn{2}{c|}{$(1,\sqrt{3}-2,1-\sqrt{3})$} \\
\hline\hline
\multicolumn{6}{|c|}{The alignments of triplet modular forms $Y_{\mathbf{3}, \mathbf{3'}}(\gamma\tau_{ST})$ of level 3 up to weight 6}\\ \hline
$\gamma$ & $\gamma\tau_{ST}$ & $Y^{(2)}_\mathbf{3}(\gamma\tau_{ST})$ & $Y^{(4)}_\mathbf{3}(\gamma\tau_{ST})$ & $Y^{(6)}_{\mathbf{3}, I}(\gamma\tau_{ST})$ & $Y^{(6)}_{\mathbf{3}, II}(\gamma\tau_{ST})$ \\\hline
$\{1,ST,T^{2}S\}$ & $\frac{-1+i\sqrt{3}}{2}$& $(1,\omega,\frac{-1}{2}\omega^{2})$ & $(1,\frac{-1}{2}\omega,\omega^{2})$ & \multirow{4}*{$(0,0,0)$} & $(1,-2\omega,-2\omega^{2})$ \\\cline{1-4}\cline{6-6}
$\{T,ST^{2}S,S\}$ & $\frac{1+i\sqrt{3}}{2}$& $(1,\omega^{2},-\frac{1}{2}\omega)$ & $(1,-\frac{1}{2}\omega^{2},\omega)$ && $(1,-2\omega^{2},-2\omega)$ \\\cline{1-4}\cline{6-6}
$\{TS,T^{2},T^{2}ST\}$ & $2+\omega$& $(1,1,-\frac{1}{2})$ & $(1,-\frac{1}{2},1)$ && $(1,-2,-2)$ \\\cline{1-4}\cline{6-6}
$\{STS,ST^{2},TST^{2}\}$ & $\frac{-3+i\sqrt{3}}{6}$ & $(0,0,1)$ & $(0,1,0)$ && $(1,0,0)$ \\\hline \hline
\multicolumn{6}{|c|}{The alignments of triplet modular forms $Y_{\mathbf{3}, \mathbf{3'}}(\gamma\tau_{TS})$ of level 3 up to weight 6}\\ \hline
$\gamma$ & $\gamma\tau_{TS}$ & $Y^{(2)}_\mathbf{3}(\gamma\tau_{TS})$ & $Y^{(4)}_\mathbf{3}(\gamma\tau_{TS})$ & $Y^{(6)}_{\mathbf{3}, I}(\gamma\tau_{TS})$ & $Y^{(6)}_{\mathbf{3}, II}(\gamma\tau_{TS})$ \\ \hline
$\{1,TS,ST^{2}\}$ & $\frac{1+i\sqrt{3}}{2}$& $(1,\omega^{2},-\frac{1}{2}\omega)$ & $(1,-\frac{1}{2}\omega^{2},\omega)$ & \multirow{4}*{$(0,0,0)$} & $(1,-2\omega^{2},-2\omega)$ \\ \cline{1-4}\cline{6-6}
$\{T,T^{2}S,TST^{2}\}$ & $\frac{3+i\sqrt{3}}{2}$ & $(1,1,-\frac{1}{2})$ & $(1,-\frac{1}{2},1)$ && $(1,-2,-2)$ \\ \cline{1-4}\cline{6-6}
  $\{ST,ST^{2}S,T^{2}ST\}$ & $\frac{(-1)^{5/6}}{\sqrt{3}}$ & $(0,0,1)$ & $(0,1,0)$ && $(1,0,0)$ \\ \cline{1-4}\cline{6-6}
  $\{STS,T^{2},S\}$ & $2+\omega$ & $(1,\omega,\frac{-1}{2}\omega^{2})$ & $(1,\frac{-1}{2}\omega,\omega^{2})$ && $(1,-2\omega,-2\omega^{2})$ \\
  \hline\hline
  \multicolumn{6}{|c|}{The alignments of triplet modular forms $Y_{\mathbf{3}, \mathbf{3'}}(\gamma\tau_{T})$ of level 3 up to weight 6}\\ \hline
$\gamma$ & $\gamma\tau_{T}$ & \multicolumn{3}{c|}{$Y^{(2)}_\mathbf{3}(\gamma\tau_{T})$, $Y^{(6)}_{\mathbf{3}, I}(\gamma\tau_{T})$, $Y^{(4)}_\mathbf{3}(\gamma\tau_{T})$}  & $Y^{(6)}_{\mathbf{3}, II}(\gamma\tau_{T})$ \\ \hline
$\{1,T,T^{2}\}$ & $i \infty$& \multicolumn{3}{c|}{$(1,0,0)$} & \multirow{4}*{$(0,0,0)$}  \\ \cline{1-5}
$\{ST,ST^{2},S\}$ & $0$& \multicolumn{3}{c|}{$(1,-2,-2)$} &   \\\cline{1-5}
$\{TS,ST^{2}S,TST^{2}\}$ & $1$& \multicolumn{3}{c|}{$(1,-2\omega,-2\omega^{2})$} &   \\ \cline{1-5}
$\{STS,T^{2}S,T^{2}ST\}$ & $-1$& \multicolumn{3}{c|}{$(1,-2\omega^{2},-2\omega)$} &  \\ \hline \hline
\end{tabular} }
\caption{\label{tab:FP&vacuum_A4}
The alignments of the triplet modular forms $Y^{(2)}_{\mathbf{3}, \mathbf{3'}}(\tau_f)$, $Y^{(4)}_{\mathbf{3}, \mathbf{3'}}(\tau_f)$  and $Y^{(6)}_{\mathbf{3}, \mathbf{3'}}(\tau_f)$ of level 3 at the fixed point $\tau_f=\gamma\tau_S$, $\gamma\tau_{ST}$, $\gamma\tau_{TS}$, $\gamma\tau_{T}$ with $\gamma \in A_4$. We have identified the modulus parameter $\tau$ with $T^3\tau=\tau+3$ in the second column because $Y_{\mathbf{r}}(\tau)=Y_{\mathbf{r}}(\tau+3)$ for any modular multiplet $Y_{\mathbf{r}}$ of level 3. }
\end{center}
\end{table}

In the framework of tri-direct modular model with two-right handed neutrinos, we have considered all possible residual symmetries in different sectors, yet no viable models can be found if the modular weights of $Y_{\mathrm{atm}}$ and $Y_{\mathrm{sol}}$ are equal to 2 or 4. Some models compatible with experimental data can be obtained if either $Y_{\mathrm{atm}}$ or $Y_{\mathrm{sol}}$ is a weight 6 modular form transforming as $\mathbf{3}$ under $A_4$ such that its alignment is not fixed uniquely. We give one example in the following, the residual symmetry of the charged lepton sector is $G_e=Z^{T}_3$ and
\begin{equation}
Y_{\mathrm{atm}}\sim Y^{(2)}_{\mathbf{3}},~~~Y_{\mathrm{sol}}\sim Y^{(6)}_{\mathbf{3}},~~~\tau_{f, \mathrm{atm}}=ST\tau_S=\frac{-1+i}{2},~~~\tau_{f,\mathrm{sol}}=\tau_S=i\,.
\end{equation}
The neutrino mass matrix is of the form of Eq.~\eqref{eq:mnu_3deg}, and the measured values of the mixing parameters and neutrino masses can be accommodated very well, e.g.
\begin{eqnarray}
\nonumber&& m_a=2.090\,\mathrm{meV},~~r_1=0.982,~~\eta_1=1.354\pi,~~r_2=0.385,~~\eta_2=1.683\pi\,,\\
\nonumber&&\sin^2\theta_{13}=0.0224,~~\sin^2\theta_{12}=0.310,~~\sin^2\theta_{23}=0.563,~~\delta_{CP}=-0.399\pi,\\
&&\beta=-0.124\pi,~~m_1=0\,\mathrm{meV},~~m_2=8.597\,\mathrm{meV},~~m_3=50.279\,\mathrm{meV}\,.
\end{eqnarray}
In the tri-direct modular model with three right-handed neutrinos, we can also find many phenomenologically viable models from $A_4$ modular symmetry. We shall not list all the possibilities but give an example, we take $G_{e}=Z^{ST}_3$ and the atmospheric, solar and ``decoupled'' modular forms are assigned to transform as
\begin{equation}
\begin{aligned}
& Y_{\mathrm{atm}}\sim Y^{(2)}_{\mathbf{3}},~~~~ Y_{\mathrm{sol}}\sim Y^{(4)}_{\mathbf{3}},~~~~Y_{\mathrm{dec}}\sim Y^{(2)}_{\mathbf{3}},\\
&\tau_{f, \mathrm{atm}}=\tau_S=i,~~\tau_{f, \mathrm{sol}}=\tau_S=i,~~\tau_{f, \mathrm{dec}}=TS\tau_{ST}=2+\omega\,.
\end{aligned}
\end{equation}
The neutrino mass matrix in this case is of the following form
\begin{eqnarray}
\nonumber m_{\nu}&=&m_a\left[\left(
\begin{array}{ccc}
1 & -2+\sqrt{3}  &  1-\sqrt{3} \\
-2+\sqrt{3}  &  7-4 \sqrt{3}   &  -5+3 \sqrt{3} \\
1-\sqrt{3}   &  -5+3 \sqrt{3}  &  4-2 \sqrt{3}
\end{array}
\right)+r_1e^{i\eta_1}\left(
\begin{array}{ccc}
 1 & 1 & 1 \\
 1 & 1 & 1 \\
 1 & 1 & 1
\end{array}
\right) \right.\\
&&\quad +\left.
r_2e^{i\eta_2}\left(
\begin{array}{ccc}
 1 & -\frac{1}{2} & 1 \\
 -\frac{1}{2} & \frac{1}{4} & -\frac{1}{2} \\
 1 & -\frac{1}{2} & 1 \\
\end{array}
\right)\right]\,.
\end{eqnarray}
The agreement with experiment data is optimised for
\begin{equation}
m_{a}=48.353\,\text{meV}, ~~ r_{1}=0.705,~~\eta_{1}=0.066\pi, ~~ r_{2}=0.859, ~~ \eta_2=1.037 \pi\,,
\end{equation}
and accordingly the neutrino masses and mixing parameters are determined to be
\begin{eqnarray}
\nonumber && \sin^2\theta_{13}=0.0224, ~~~  \sin^2\theta_{12}=0.310, ~~~ \sin^2\theta_{23}=0.563\,,\\
\nonumber &&  \delta_{CP}=-0.353\pi, ~~~ \alpha_{21}=-0.676\pi, ~~~ \alpha_{31}=-1.551\pi \,, \\
&& m_1=72.207\,\text{meV}, ~~~m_2=72.717\,\text{meV}, ~~~ m_3= 87.988\,\text{meV}\,.
\end{eqnarray}

\section{\label{sec:conclusion} Conclusion }

Models based on modular flavour symmetry are highly predictive. In the most economical version of modular symmetry models, the VEV of the modulus $\tau$ is the unique source of the flavour symmetry breaking, and the flavon fields are not absolutely necessary. If the VEV of $\tau$ takes some special values, certain residual subgroup of the modular symmetry would be preserved. Inside the fundamental domain, we have derived the four nontrivial fixed points $\tau_S=i$, $\tau_{ST}=-\frac{1}{2}+i\frac{\sqrt{3}}{2}$, $\tau_{TS}=\frac{1}{2}+i\frac{\sqrt{3}}{2}$, $\tau_T=i\infty$ which
are invariant under the actions of the modular subgroups $Z^{S}_2$, $Z^{ST}_3$, $Z^{TS}_3$ and $Z^{T}_N$ respectively with $N$ being the level of modular group. Other fixed points of modulus are related to $\tau_S$, $\tau_{ST}$, $\tau_{TS}$ and $\tau_T=i\infty$ through modular transformation. The most general form of the fixed modulus is given by $\tau_f=\gamma\tau_S$, $\gamma\tau_{ST}$, $\gamma\tau_{TS}$, $\gamma\tau_{T}$ with $\gamma\in\overline{\Gamma}$. Although the number of fixed point $\tau_f$ is infinity, the inequivalent directions of the modular forms $Y_\mathbf{r}(\tau_f)$ at fixed points are finite. We have shown that it is sufficient to only consider the values of  $Y_\mathbf{r}(\tau_f)$ with $\tau_f$ given by
$\Gamma_N\tau_S$, $\Gamma_N\tau_{ST}$, $\Gamma_N\tau_{TS}$ and $\Gamma_N\tau_{T}$.

Before applying these results to lepton mixing,
we first proved that it is necessary for the neutrino and charged lepton sectors to have different residual symmetry, since if they
shared a common symmetry value of $\tau_f$, the lepton mixing would contain four or six zeros which is excluded by present data.
The approach to lepton mixing followed here is motivated by the tri-direct CP approach in usual discrete flavour symmetry, however here we find that it is unnecessary to CP, as we discuss below. For example, in the minimal scenario with two right-handed neutrinos, we assumed that the modular symmetry is broken to different residual subgroups in the charged lepton sector, atmospheric neutrino sector and solar neutrino sector. In this case, the two columns of the Dirac neutrino mass matrix are triplet modular forms whose alignments are determined by residual symmetry at fixed points $\tau_f$.

Focussing on level $N=4$, corresponding to the flavour group $S_4$,
we considered all the resulting triplet modular forms at the above fixed points up to weight 6.
We then applied the results to lepton mixing, with different residual subgroups in the charged lepton sector and each of the right-handed neutrinos sectors. In the minimal case of two right-handed neutrinos,
we found three phenomenologically viable cases, denoted A, B and C.
We showed that if only weight 2 and weight 4 modular forms are involved in the neutrino Yukawa coupling, the light neutrino mass matrix involves three real parameters and three independent cases can be compatible with the experimental data, as summarized in table~\ref{tab:assign_TDM}. The cases A and B lead to TM1 mixing pattern, and the first column of the lepton mixing matrix is approximately $(0.800, 0.421, 0.428)^{T}$ for the case C. The case B corresponds to the CSD($n$) model with
$n=1+\sqrt{6}\approx 3.45$, intermediate between CSD$(3)$ and CSD$(4)$.

We find it remarkable that, even without CP,  the modular symmetry uniquely fixes the value of $n$ to be a unique real number
(to be compared to the usual tri-direct approach, where $n$ takes an arbitrary real value, assuming a CP symmetry). If a higher weight (e.g., weight 6) modular form enters in the neutrino Yukawa coupling, its alignment is not constant vector any more at fixed point $\tau_f$ and more free parameters are involved in the light neutrino mass matrix. Hence we find many cases can described the measured values of lepton masses and mixing in this scenario, and we give one example for illustration. Furthermore, we generalized the tri-direct modular approach to the models with three right-handed neutrinos, and many cases in agreement with experimental data can also be found.

In the same fashion, we also studied the possible tri-direct modular models for the level $N=3$ case, corresponding to $A_4$ flavour symmetry. In the minimal model with two right-handed neutrinos, no models can be compatible with data if the modular weights of $Y_{\mathrm{atm}}$ and $Y_{\mathrm{sol}}$ are equal to 2 or 4. However, phenomenologically viable models can be found if the modular weights of either $Y_{\mathrm{atm}}$ or $Y_{\mathrm{sol}}$ are greater than 4 or three right-handed neutrinos are involved.

In the tri-direct modular approach here, the physical results only depend on the structure of modular symmetry group and the assumed residual symmetries. The details of the mechanism realising the desired residual symmetries are irrelevant. It seems that at least three complex moduli $\tau_e$, $\tau_{\mathrm{atm}}$ and $\tau_{\mathrm{sol}}$ responsible for the breaking of modular symmetry in the charged lepton sector, atmospheric neutrino sector and solar neutrino sector may be needed. It would be interesting to implement one of the successful cases A, B, C in a concrete model based on multiple modular symmetries~\cite{deMedeirosVarzielas:2019cyj,King:2019vhv} or some other mechanism which can lead to different values of $\tau$ in the charged lepton, atmospheric neutrino and solar neutrino sectors, and this is left for future work.

\section*{Acknowledgements}
G.-J.\, D., X.-G.\, L. and J.-N.\, L. acknowledges the support of the National Natural Science Foundation of China under Grant Nos
11975224 and 11835013. S.\,F.\,K. acknowledges the STFC Consolidated Grant ST/L000296/1 and the European Union's Horizon 2020 research and innovation program under the Marie Sk\l{}odowska-Curie grant agreements Elusives ITN No.\ 674896 and InvisiblesPlus RISE No.\ 690575. S.\,F.\,K. also acknowledges the hospitality of G.-J.\, D. during this collaboration.
We acknowledge Prof. Shun Zhou and Dr. Newton Nath for stimulating discussion in the initial stage of this work.

\section*{Appendix}

\setcounter{equation}{0}
\renewcommand{\theequation}{\thesection.\arabic{equation}}
%%%%%%%%%%%%%%%%%%%%%%%%%%%%%%%%%%%%%%%%%%
\begin{appendix}
%%%%%%%%%%%%%%%%%%%%%%%%%%%%%%%%%%%%%%%%%%
\section{\label{sec:S4_group_app}Group theory of $\Gamma_4\cong S_{4}$}

The finite modular group $\Gamma_4$ has two generators $S$ and $T$ which fulfill the following  rations
\begin{equation}
S^2=(ST)^3=(TS)^3=T^4=1\,.
\end{equation}
The finite modular group $\Gamma_4$ is isomorphic to the permutation group $S_4$ of four objects. In order to see the correlation between $S_4$ and tri-bimaximal mixing and the connection to $S_3$, $A_4$ groups more easily, it is convenient to generate the $S_4$ group in terms of three generators $\hat{S}$, $\hat{T}$ and $\hat{U}$ with the multiplication rules~\cite{Hagedorn:2010th,Ding:2013hpa},
\begin{eqnarray}
\hat{S}^2=\hat{T}^3=\hat{U}^2=(\hat{S}\hat{T})^3=(\hat{S}\hat{U})^2=(\hat{T}\hat{U})^2=(\hat{S}\hat{T}\hat{U})^4=1\,,
\end{eqnarray}
where $\hat{S}$ and $\hat{T}$ alone generate the group $A_4$, while $\hat{T}$ and $\hat{U}$ alone generate the group $S_3$. The generators $S$, $T$ can be expressed in terms of $\hat{S}$, $\hat{T}$ and $\hat{U}$ \begin{equation}
S = \hat{S}\hat{U},\, T = \hat{S} \hat{T}^2 \hat{U},\,ST = \hat{T}
\end{equation}
or vice versa
\begin{equation}
\hat{S}=(ST^2) ^2,~~~ \hat{T}=ST,~~~\hat{U}=T^2ST^2\,.
\end{equation}
The $S_4$ group has five conjugacy classes as follow,
\begin{eqnarray}
\nonumber 1C_1 &=& \left\{1 \right\}, \\
\nonumber 3C_2 &=& \left\{(ST^2)^2, T^2, ST^2S \right\},  \\
\nonumber 6C_2^{\prime} &=& \left\{T^2ST^2,TST^3,S,TST^2S,T^3ST,ST^2ST \right\},  \\
\nonumber 8C_3 &=& \left\{ST,T^2ST^3,T^2ST,(TS)^2,(ST)^2,T^3ST^2,TST^2,TS\right\},  \\
6C_4 &=& \left\{TST,STS,T^3,T,ST^2,T^2S\right\}\,,
\end{eqnarray}
where $nC_k$ denotes a conjugacy class with $n$ elements and the  subscript $k$ is the order of the elements. The group $S_4$ has five irreducible representations: two singlets $\mathbf{1}$ and $\mathbf{1}^{\prime}$, one doublet $\mathbf{2}$ and two triplets $\mathbf{3}$ and $\mathbf{3}^{\prime}$. In the present work, we choose the same basis as that of~\cite{Ding:2013hpa}. The explicit forms of $S$ and $T$ in each of the irreducible representations are summarized in table~\ref{tab:S4_rep_hat}. Notice that the triplet representations $\mathbf{3}$ and $\mathbf{3}^{\prime}$ correspond to $\mathbf{3}^{\prime}$ and $\mathbf{3}$ of Refs.~\cite{Penedo:2018nmg,Novichkov:2018ovf} respectively.

\begin{table}[t!]
\begin{center}
\begin{tabular}{|c|c|c|}\hline\hline
 ~~  &  $S$  &   $T$     \\ \hline
~~~$\mathbf{1}$, $\mathbf{1}^\prime$ ~~~ & $\pm1$   &  $\pm1$  \\ \hline
   &   &       \\ [-0.16in]
$\mathbf{2}$ &  $\left( \begin{array}{cc}
 0&~1 \\
 1&~0
 \end{array} \right) $
 & $\left( \begin{array}{cc}
 0 ~&~ \omega^2 \\
 \omega ~&~ 0
\end{array} \right) $
    \\ [0.15in]\hline
   &   &       \\ [-0.16in]
$\mathbf{3}$, $\mathbf{3}^\prime $ & $\pm\frac{1}{3} \left(\begin{array}{ccc}
 1 ~& -2  &~ -2  \\
 -2  ~& -2  &~ 1 \\
 -2  ~& 1 &~ -2
\end{array}\right)$ &
$\pm\frac{1}{3}\left( \begin{array}{ccc}
1 ~& -2\omega^2 &~ -2\omega \\
-2 ~& -2\omega^2 &~ \omega \\
-2  ~& \omega^2 &~ -2\omega
\end{array}\right) $
\\[0.25in] \hline\hline
\end{tabular}
\caption{\label{tab:S4_rep_hat}The representation matrices of the generators $S$ and $T$ in the five irreducible representations of $S_4$, where $\omega=e^{2\pi i/3}=-1/2+i\sqrt{3}/2$ is a cubic root of unity. }
\end{center}
\end{table}

The Clebsch-Gordan coefficients in our working basis can be found in~\cite{Ding:2013hpa}, for completeness we present them in the following. We shall use $\alpha_i$ to denote the elements of first representation and $\beta_i$ stands for the elements of the second representation of the tensor product.
\begin{eqnarray}
\mathbf{1}^\prime\otimes\mathbf{2}&=&\mathbf{2}\sim
\left(\begin{array}{c}\alpha \beta_1 \\-\alpha \beta_2\end{array}\right)\,,\\
\mathbf{1}^\prime\otimes\mathbf{3}&=&\mathbf{3}^\prime\sim
\left(\begin{array}{c}\alpha \beta_1  \\ \alpha \beta_2  \\ \alpha \beta_3 \end{array}\right)\,,\\
\mathbf{1}^\prime\otimes\mathbf{3}^\prime&=&\mathbf{3}\sim
\left(\begin{array}{c} \alpha \beta_1  \\ \alpha \beta_2  \\ \alpha \beta_3 \end{array}\right)\,.
\end{eqnarray}

\begin{eqnarray}
\mathbf{2}\otimes\mathbf{2}=\mathbf{1}\oplus\mathbf{1^\prime}\oplus\mathbf{2}~~~&\mathrm{with}&~~~\left\{
\begin{array}{l}
\mathbf{1}\sim \alpha_1 \beta_2+ \alpha_2 \beta_1\\ [0.1in]
\mathbf{1}^\prime\sim \alpha_1 \beta_2- \alpha_2 \beta_1\\ [0.1in]
\mathbf{2}\sim
\left(\begin{array}{c} \alpha_2 \beta_2  \\  \alpha_1 \beta_1 \end{array}\right)
\end{array}
\right. \\
\mathbf{2}\otimes\mathbf{3}=\mathbf{3}\oplus\mathbf{3}^\prime~~~&\mathrm{with}&~~~\left\{
\begin{array}{l}
\mathbf{3}\sim
\left(\begin{array}{c}  \alpha_1 \beta_2+ \alpha_2 \beta_3 \\  \alpha_1 \beta_3+ \alpha_2 \beta_1  \\  \alpha_1 \beta_1+ \alpha_2 \beta_2 \end{array}\right) \\ \\[-0.1in]
\mathbf{3}^\prime\sim
\left(\begin{array}{c} \alpha_1 \beta_2- \alpha_2 \beta_3 \\  \alpha_1 \beta_3- \alpha_2 \beta_1  \\  \alpha_1 \beta_1- \alpha_2 \beta_2 \end{array}\right)
\end{array}
\right. \\
\mathbf{2}\otimes\mathbf{3^\prime}=\mathbf{3}\oplus\mathbf{3^\prime}~~~&\mathrm{with}&~~~\left\{
\begin{array}{l}
\mathbf{3}\sim
\left(\begin{array}{c}  \alpha_1 \beta_2- \alpha_2 \beta_3 \\  \alpha_1 \beta_3- \alpha_2 \beta_1  \\  \alpha_1 \beta_1- \alpha_2 \beta_2 \end{array}\right)\\ \\[-0.1in]
\mathbf{3}^\prime\sim
\left(\begin{array}{c} \alpha_1 \beta_2+ \alpha_2 \beta_3 \\  \alpha_1 \beta_3+ \alpha_2 \beta_1  \\  \alpha_1 \beta_1+ \alpha_2 \beta_2\end{array}\right)
\end{array}
\right. \\
\mathbf{3}\otimes\mathbf{3}=\mathbf{3}^\prime\otimes\mathbf{3}^\prime=\mathbf{1}\oplus\mathbf{2}\oplus\mathbf{3}\oplus\mathbf{3}^\prime~~~&\mathrm{with}&~~~\left\{
\begin{array}{l}
\mathbf{1}\sim \alpha_1\beta_1+\alpha_2\beta_3+ \alpha_3 \beta_2 \\ [0.1in]
\mathbf{2}\sim
\left(\begin{array}{c} \alpha_2 \beta_2+ \alpha_1 \beta_3+ \alpha_3 \beta_1  \\  \alpha_3 \beta_3+ \alpha_1 \beta_2+ \alpha_2 \beta_1 \end{array}\right) \\ \\[-0.1in]
\mathbf{3}\sim
\left(\begin{array}{c} \alpha_2 \beta_3- \alpha_3 \beta_2  \\  \alpha_1 \beta_2- \alpha_2 \beta_1  \\  \alpha_3 \beta_1- \alpha_1 \beta_3 \end{array}\right) \\ \\[-0.1in]
\mathbf{3}^\prime\sim
\left(\begin{array}{c} 2 \alpha_1 \beta_1- \alpha_2 \beta_3- \alpha_3 \beta_2  \\  2 \alpha_3 \beta_3- \alpha_1 \beta_2- \alpha_2 \beta_1  \\
2 \alpha_2 \beta_2-\alpha_1\beta_3-\alpha_3\beta_1 \end{array}\right)
\end{array}
\right. \\
\mathbf{3}\otimes\mathbf{3}^\prime=\mathbf{1}^\prime\oplus\mathbf{2}\oplus\mathbf{3}\oplus\mathbf{3}^\prime~~~&\mathrm{with}&~~~\left\{
\begin{array}{l}
\mathbf{1}^\prime\sim \alpha_1\beta_1+\alpha_2\beta_3+\alpha_3\beta_2\\ [0.1in]
\mathbf{2}\sim
\left(\begin{array}{c} \alpha_2 \beta_2+ \alpha_1 \beta_3+ \alpha_3 \beta_1  \\ -( \alpha_3 \beta_3+ \alpha_1 \beta_2+ \alpha_2 \beta_1) \end{array}\right) \\ \\[-0.1in]
\mathbf{3}\sim
\left(\begin{array}{c}2 \alpha_1 \beta_1-  \alpha_2 \beta_3- \alpha_3 \beta_2  \\ 2 \alpha_3 \beta_3-  \alpha_1 \beta_2- \alpha_2 \beta_1  \\
2\alpha_2 \beta_2-\alpha_1\beta_3-\alpha_3\beta_1 \end{array}\right) \\ \\[-0.1in]
\mathbf{3}^\prime\sim
\left(\begin{array}{c} \alpha_2 \beta_3- \alpha_3 \beta_2  \\  \alpha_1 \beta_2- \alpha_2 \beta_1  \\  \alpha_3 \beta_1- \alpha_1 \beta_3 \end{array}\right)
\end{array}
\right.
\end{eqnarray}

\end{appendix}

\newpage

\providecommand{\href}[2]{#2}\begingroup\raggedright\endgroup

\end{document}